\documentclass[prb,aps,amssym,nofootinbib,floatfix,twocolumn]{revtex4} 
\usepackage{epsfig,natbib}
\newcommand{\be}{\begin{equation}}
\newcommand{\ee}{\end{equation}}
\newcommand{\bea}{\begin{eqnarray}}
\newcommand{\eea}{\end{eqnarray}}
\newcommand{\ket}{\rangle}
\newcommand{\bra}{\langle}

\begin{document}
\title{Spectral Function for the S=1 Heisenberg Antiferromagetic Chain}
\author{Steven R. White}
\affiliation{Department of Physics and Astronomy, University of California,
 Irvine CA 92697, USA}
\author{Ian Affleck}
\affiliation{Department of Physics and Astronomy, University of British
  Columbia, Vancouver, British Columbia, Canada V6T 1Z1}
\date{today}
\begin{abstract}
We study the spectral function, $S(k,\omega )$ for the spin-1, one
dimensional antiferromagnetic chain using a time-dependent density
matrix renormalizaton group (DMRG) numerical method. We develop
methods for extrapolating the time dependent correlation functions
to larger times in order to enhance the frequency resolution.  The
resulting spectral functions are impressively precise and accurate.
Our results confirm many qualitative expectations from non-linear
$\sigma$ model methods and test them quantitatively. The critical
wave-vector at which the single particle excitation emerges from the
2-particle continuum is estimated to be $0.23\pi-0.24\pi$. 
\end{abstract}
\maketitle
\section{Introduction}
The $S=1$ Heisenberg antiferromagnetic chain,
\be H=J\sum_j\vec S_j\cdot \vec S_{j+1},\ee
has been a subject of intense theoretical and experimental study since 
Haldane's observation that it has an excitation gap above a singlet 
ground state to a triplet excited state, quite unlike the S=1/2 case. Much has been learned
about the time independent properties using a combination of analytic and numerical methods,
in particular the density matrix renormalization group (DMRG).\cite{dmrg}  Our knowledge of dynamical
properties, while substantial, leaves more room for improvement.\cite{kuhner}
Recently, DMRG has been
significantly extended to allow direct calculation of time dependence (tDMRG)\cite{vidal-time,whitefeiguin,daley,feiguinwhite} and 
time-dependent correlation
functions in particular. In this paper we will utilize tDMRG to calculate high-resolution spectral
functions for the $S=1$ chain for a broad range of momenta.  Some of the numerical techniques developed here were
briefly described in Ref. \onlinecite{pereira}.

The spectral function we shall focus on, in the isotropic case, is
\be  S(k,\omega )=\sum_{j=-\infty}^\infty e^{-ikj}\int_{-\infty}^\infty dt \ e^{i\omega t}
\bra 0|S^a_j(t)S^a_{0}(0)|0\ket \label{S}\ee
where 
$S(k,\omega) \equiv S^{aa}(k,\omega )$, with $a = x$, $y$, or $z$, and the subscripts
on $S$ indicate sites.  The tDMRG method
calculates the space-time dependent expectation values appearing in Eq. (\ref{S}) directly,
and then one performs  the Fourier transforms (FTs) in Eq. (\ref{S}) 
to obtain $S(k,\omega )$.  The crucial practical
issue for this approach, which we discuss in detail in this paper, is dealing with the finite range
of times available from a tDMRG simulation.  We find that two different methods for extrapolation in time to increase the
range of times used in the FTs  are both very useful for increasing the frequency resolution.

\begin{figure}[htbp]
\begin{center}
\includegraphics*[width=0.6\hsize,scale=1.0]{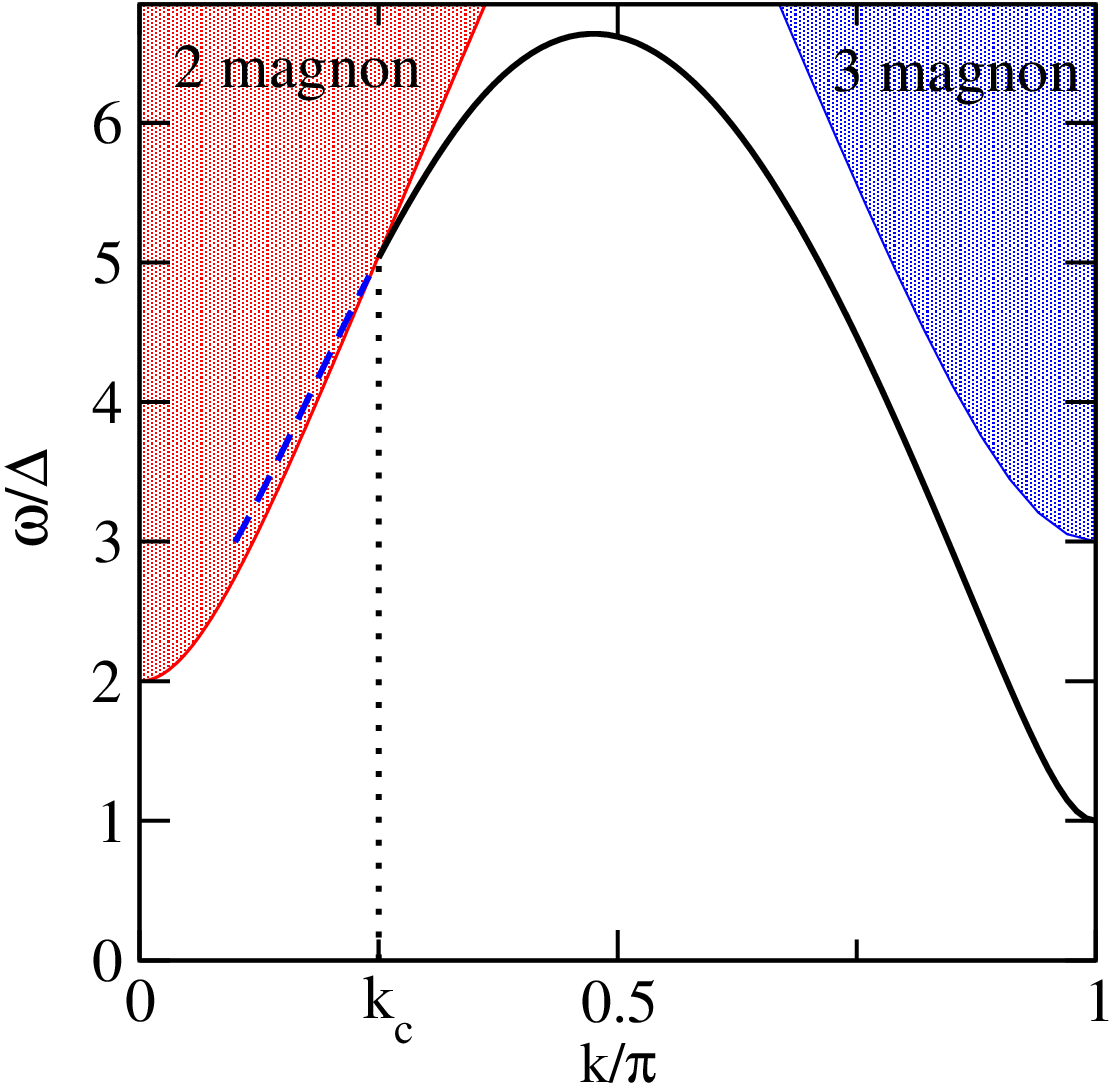}
\caption{General features of the spectral function. The four and higher magnon continua are not shown.
}
\label{fig:spec_wt}
\end{center}
\end{figure}

Much insight into the model can be obtained from an approximate mapping, 
based on the large S limit,
onto the non-linear $\sigma$-model (NL$\sigma$M) with Lagrangian density:
\be \mathcal{L}={1\over 2gv}[(\partial_t\vec \phi )^2-v^2\partial_x\vec \phi )^2],\ee
where the lattice spin operators are represented as:
\be \vec S_j\approx {1\over vg}\vec \phi \times \partial_t\vec \phi +(-1)^js\vec \phi ,\ee
$\phi^2=1$, $v$ is the spin velocity and $g\approx 2/s$ is the coupling constant. $\vec \phi $
is the antiferromagnetic order parameter and $\vec l\equiv (1/vg)\vec \phi \times \partial_0\vec \phi$
is the uniform magnetization density. This low energy representation is only 
valid for wave-vectors near $0$ and $\pi$. This field theory is known 
to have a singlet ground state in one space dimension, with a gap, $\Delta$, to 
a massive spin-triplet of excitations.  There are no bound states 
or any other single particle excitations besides this triplet.  This 
implies that a single particle ($\delta$-function) peak should appear in $S(k,\omega)$
near $k=\pi$ at energy $\Delta$ but near $k\approx 0$ the lowest excitations are a 2-particle continuum
starting at $2\Delta$. 
Interaction effects mix $\vec \phi$ with $(\vec \phi )^3$ [but not $(\vec \phi )^2$ 
by symmetry] so that the theory 
also predicts a 3-particle continuum at $k\approx \pi$ starting at $3\Delta$, a 4-particle 
continuum at $k\approx 0$, et cetera.  
Continuity between $k\approx 0$ and $k\approx \pi$ then implies a 
qualitative sketch like Fig. \ref{fig:spec_wt} for the region of non-zero spectral weight.  In particular, 
the single particle peak must emerge out of the 2-particle continuum at some critical 
wave-vector $k_c$.  Integrability of the NL$\sigma$M allows for the 
calculation of exact form factors and hence predictions for the 
detailed shape of the 2-particle contribution to $S(k,\omega )$ near $k\approx 0$ 
and the 3-particle contribution near $k\approx \pi$.\cite{weston,horton}

Since the field theory 
is based on $S\to \infty$ and is only valid for $k$ very close to $0$ and $\pi$ 
it is unclear how well any of these predictions should describe the S=1 case. 
Various comparisons of NL$\sigma$M predictions with numerical results 
have been performed before, but these have necessarily focused primarily on equal time 
correlators and thermodynamic quantities.  Experiments 
on quasi-1-dimensional antiferromagnets have clearly confirmed 
the Haldane gap but the 2-particle nature of the small $k$ excitations 
and the existence of a gap at $k=\pi$ from the single particle 
excitation at $\Delta$ to the bottom of the 3-particle continuum at $3\Delta$ 
have not been confirmed and have led to some questioning of the validity 
of these field theory predictions. 

One of the purposes of this 
paper is to compare our tDMRG calculations of $S(k,\omega )$  with the NL$\sigma$M predictions.
In addition, we will
determine $k_c$ and examine the behavior near this wave-vector. 

In Sec. II we review the time dependent DMRG algorithm and discuss methods for
extrapolating the time-dependent correlation function to longer times before Fourier transforming.
In Sec. III we discuss properties of the single magnon 
excitation.  In Sec. IV we discuss $S$ near $k\approx 0$ and compare with 
the  2-magnon NL$\sigma$M predictions.  In Sec. V
we study $S$ near $k\approx \pi$ and compare with 
the 1-magnon plus 3-magnons predictions.  In Sec. VI 
we study intermediate $k$ in the vicinity of $k_c$ 
where the single magnon peak first emerges. 
Sec. VII contains our conclusions. 

\section{Time dependent DMRG and extrapolation in time}
The time dependent DMRG method\cite{whitefeiguin,daley,vidal-time} has been described elsewhere; here
we describe some of the practical issues related to obtaining dynamical spectra.  First,
we describe the specific tDMRG algorithms used; second, time step and truncation errors;
and third,  extrapolation in
time and windowing for time Fourier transforms.

The first step of a calculation is to use ordinary DMRG to find the ground state of a finite open system to high
accuracy, with  typical lengths being $L=200-400$.  
To avoid $S=1/2$ end states, we put real $S=1/2$ spins on the first and last
sites. Let $i$ be one of the two center sites. After obtaining the ground state $\phi$,
we apply  the operator $S^+_i$ or $S^z_i$ to create a state which
is a mixture of excited states, $\psi(t=0)$.
We target and time evolve  $\psi$, and also, in order to minimize time step errors in
correlation functions, $\phi$.  As the time evolution occurs, we measure the correlation function
$S(j-i,t)$ when the tDMRG step is at site $j$, accumulating the function for a wide range of separations and times.\cite{whitefeiguin}

For the time evolution we use only the  Suzuki-Trotter  decomposition methods, 
which are very efficient for
chain systems with only nearest neighbor interactions.  We use two different variations.  
First, we use the original method of White
and Feiguin,\cite{whitefeiguin} which has second order Trotter errors.   An advantage of
this method is that every DMRG step has a time evolution bond operator applied, whereas Trotter 
methods splitting the links into even and odd groups\cite{daley}
 apply a time evolution operator on
only half the steps of a sweep, resulting in roughly twice as much total truncation error per
unit time evolved.  To obtain a measurement for a specific time, we perform the measurements
in a half sweep without any time evolution. Thus,
including measurements, we evolve in the repeated 6 half-sweep pattern: 
evolve left-to-right, evolve right-to-left, 
left-to-right measurement half-sweep (without time evolution), evolve right-to-left, evolve left-to-right, right-to-left measurement half-sweep.  During each half-sweep time evolution step, we
evolve a time step $\tau$, so that measurements are available with a time step of $2 \tau$.
Typically we use $\tau=0.1$. (We set $J=1$.)

The second method we use is a fourth order Trotter method,\cite{forestruth,feiguinwhite}
which with $\tau=0.1$ virtually eliminates time-step errors, at the expense of more sweeps for a given time, and 
consequently larger accumulated truncation error. 
We show below that the {\it second order} Trotter errors from the first method with a time step of  $\tau=0.1$
primarily lead to modest frequency shifts, shifting the Haldane gap by about 1\%, for example. 
(The rate of decay in time of the correlation functions seems
not to be strongly effected.)  For high accuracy studies it is more efficient to use the
fourth order method rather than simply reducing $\tau$.
The decomposition we use is\cite{forestruth}
\bea 
&e^{(A+B)\tau + {\cal O}(\tau^5)} = 
e^{A \theta \frac{\tau}{2}}
e^{B \theta \tau}
e^{A (1-\theta) \frac{\tau}{2}}
e^{B (1-2\theta) \tau}\nonumber \\
&\times e^{A (1-\theta) \frac{\tau}{2}}
e^{B \theta \tau}
e^{A \theta \frac{\tau}{2}}
\eea
with $\theta=1/(2-2^{1/3}) \approx 1.35$. Here $A$ would represent, say, the odd bonds,
and $B$ the even, and each of the seven terms is applied in a half-sweep.  Adding to this
a measurement half-sweep, a total of 
eight half-sweeps are needed to evolve by $\tau$, with 
measurements available with a time step of $\tau$.

\newcommand{\tm}{t_{\rm max}}

Generally we specify a desired truncation error for each step, and
vary the number of states kept $m$ to achieve this truncation error. However, we also
constrain $m$ to be no larger than a specified $m_{\rm limit}$ (typically 1000-2000), 
and no smaller than
a minimum $m_{\rm min}$ (typically 100-150). The purpose of $m_{\rm min}$ is to
reduce the truncation error to near zero, at little cost, in steps where the state has very small
entanglement. (These small entanglement steps occur outside the ``light cone'' of the initial disturbance
at site $i$, and time 0.) We specify a $m_{\rm limit}$ to avoid memory limitations and to avoid a 
handful of steps (near site $i$) taking a large fraction of the computer time.

\begin{figure}[htbp]
\begin{center}
\includegraphics*[width=0.7\hsize,scale=1.0]{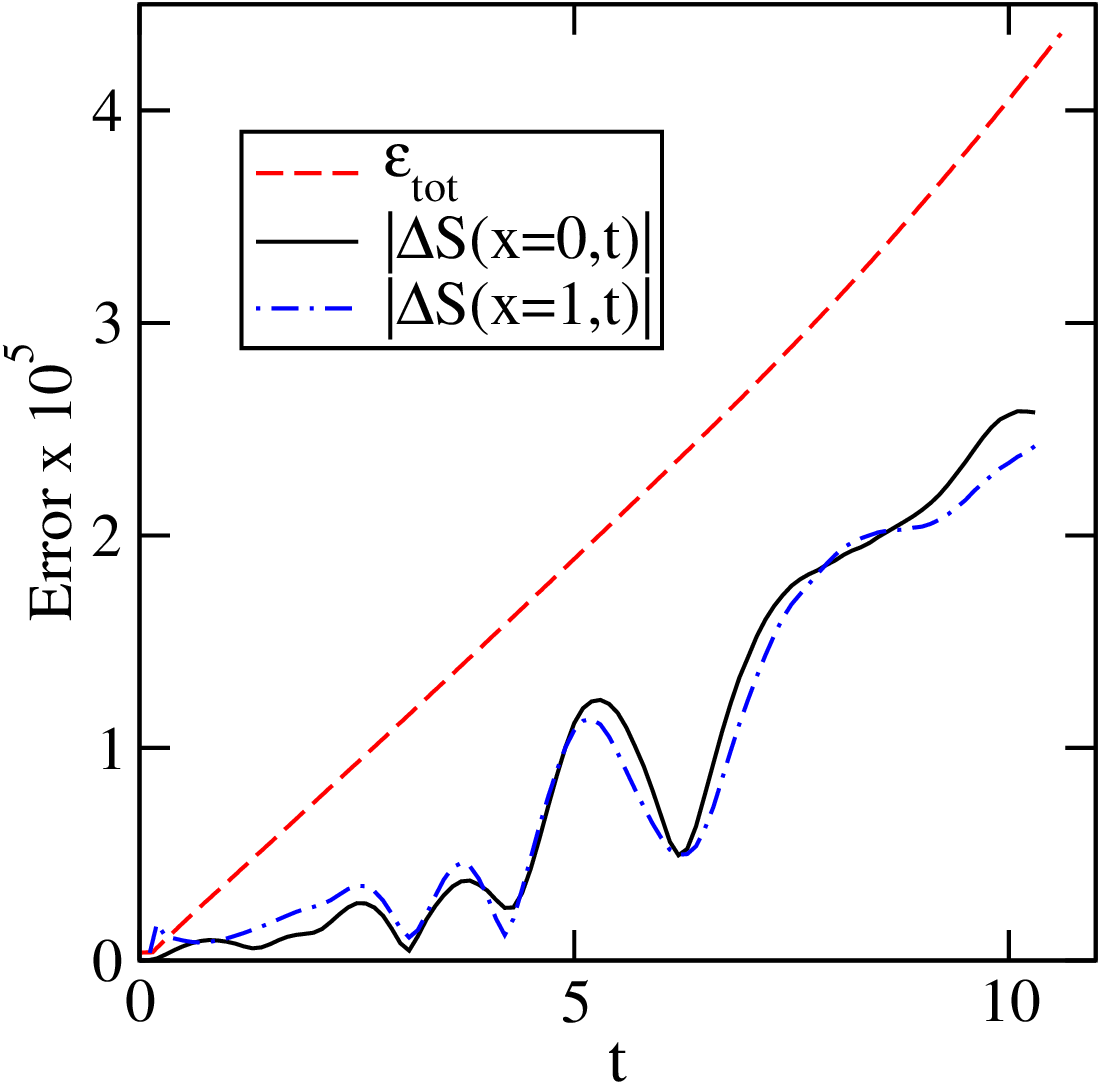}
\caption{Comparison of the errors in measurements of $S(x,t)$ and the total
accumulated truncation error $\varepsilon_{\rm tot}$. Two runs were made using the
4th order Trotter decomposition, both with $\tau=0.1$, with $\varepsilon=2\times10^{-10}$ and $\varepsilon=4\times10^{-10}$.  The values of $\Delta S$ were the difference in the results
between these
two runs. The value for $\varepsilon_{\rm tot}$ was for the larger $\varepsilon$ run.
}
\label{truncerrors}
\end{center}
\end{figure}

The total accumulated truncation error, $\varepsilon_{\rm tot}$, 
summed over all DMRG steps since the start
of the time evolution, is readily available and useful.  At each step, a small part of each wavefunction
is discarded; the truncation error is the magnitude squared of these small parts.   In 
Fig. \ref{truncerrors} we show that
 $\varepsilon_{\rm tot}$ gives a rough estimate of the 
typical errors to be expected in measurements at that time.
The plot compares $\varepsilon_{\rm tot}$ with the errors in 
$S(x,t)$ for $x=j-i=0$ and $x=1$.  The errors in $S(x,t)$ were estimated
by running with two different truncation error parameters $\varepsilon$, 
$\varepsilon=2\times10^{-10}$ and $\varepsilon=4\times10^{-10}$. We also
see that errors
of order $10^{-5}$ are computationally feasible for times $t<10$, but that the errors steadily
grow with time. In this case the errors grow roughly linearly with time because
a target truncation error per step was specified, with the upper limit $m_{\rm limit}$ playing
a small role. 

\begin{figure}[htbp]
\begin{center}
\includegraphics*[width=0.8\hsize,scale=1.0]{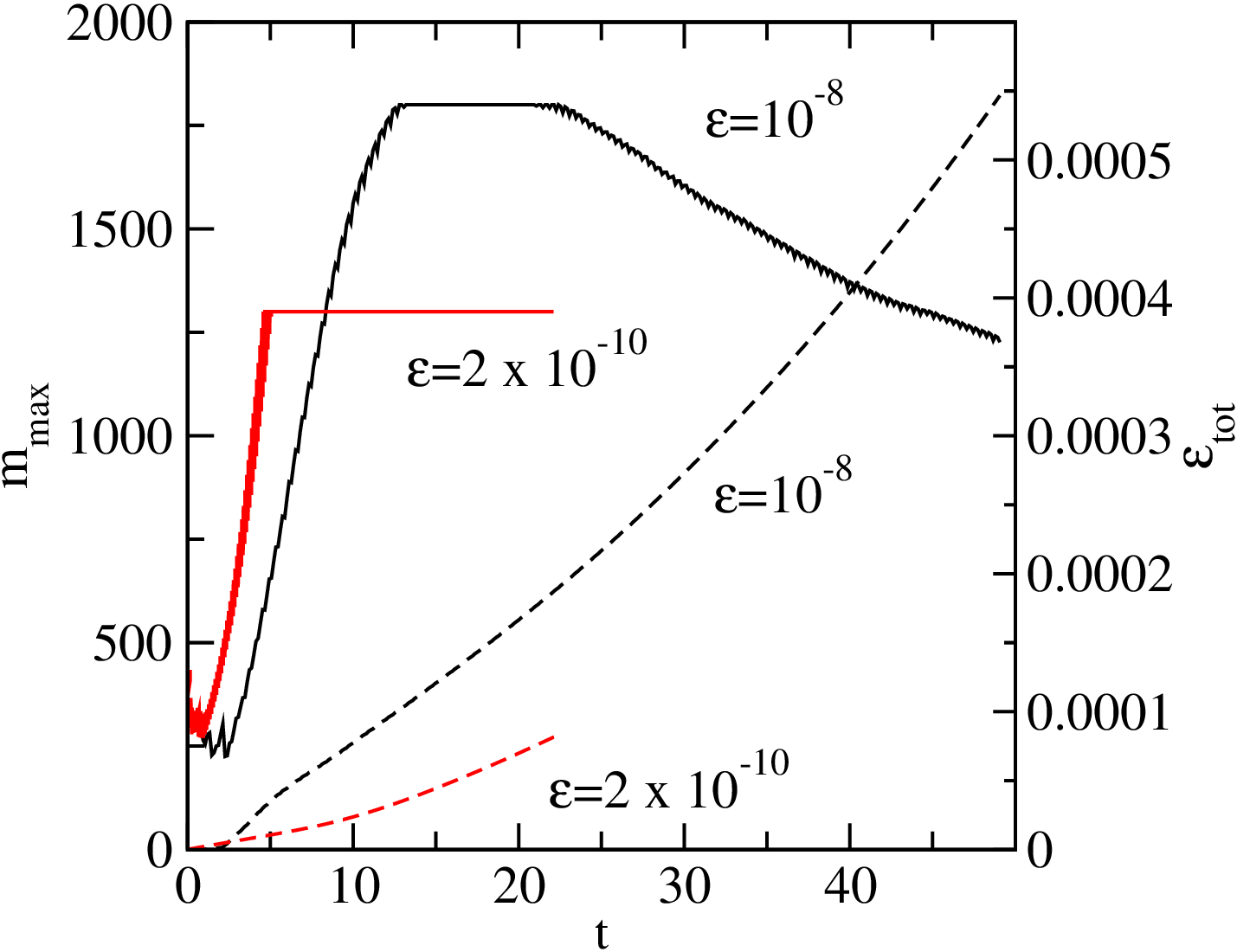}
\caption{Maximum number of states kept $m_{\rm max}$ 
versus time (upper-left curves), and total discarded weight 
$\varepsilon_{\rm tot}$ versus time (lower right curves)
 for the systems of Fig. \ref{Nwtau}. 
The simulation adjusted $m$ at each step  to try to achieve a total discarded weight of $\varepsilon$ for that step, subject to a maximum $m$ of either 1300 or 1800 (flat portions).
The figure
shows that in this case $\varepsilon_{\rm tot}$ depends more strongly on $\varepsilon$ than on the maximum
allowed $m$.  }
\label{maxmplot}
\end{center}
\end{figure}

To clarify further how this works in Fig. \ref{maxmplot} we compare 
two runs, one a second order Trotter method, the other
the more accurate run of Fig.  \ref{truncerrors}. Both runs took comparable amounts of
computer time. We define $m_{\rm max}$  to be the
maximum number of states kept over all the steps in a half-sweep. Clearly 
$m_{\rm max} \le m_{\rm limit}$.  In each half-sweep, the largest values of $m$ were
for steps near the center of the
 system,  where the spin operator was applied. After a moderate time, for these steps
 near the center $m$ was limited by $m_{\rm limit}$.  The relatively small number of
 these steps made the effect of $m_{\rm limit}$  on $\varepsilon_{\rm tot}$ small.
 The calculation time for a step is proportional to $m^3$, so limiting $m$  is important
 for efficiency.

The tDMRG yields directly the space-time dependent correlation function directly. The signal initiated by
the application of the spin operator in the center of the system spreads out with time.  Because the
system has a finite correlation length, the correlation function is only non negligible within a range
of about $|v t|$ of the center, where $v$ is the maximum spin velocity.
We always keep
the maximum time $\tm$ of the simulation small enough so that the signal has not reached the edges at
the end of the simulation.  Hence there are essentially no spatial finite size effects. One can spatially Fourier
transform (FT) $x\to k$  for any $k$; the available values of $k$ are continuous, not discrete. 
On the other hand, the feasible $\tm$ is strictly limited by the available computer time, 
and the correlation functions decay
slowly or not at all in time. 
Truncation of the signal following by an FT $t \to \omega$ would result in severe ``ringing''. The
standard approach is to multiply the signal by a windowing function, 
which typically resembles a Gaussian centered at $t=0$ but which
vanishes exactly at $\pm \tm$.  A drawback of this approach is that most of the 
data gets ``thrown away'', and the
frequency broadening of the spectrum is large. 

To avoid this over-broadening, we have developed an alternative approach based 
on linear prediction.\cite{linearprediction} 
Before the time-frequency FT, we extrapolate the time signal to long times using 
linear prediction. We then
apply a broad window which does not throw away a significant amount of the original data.
Linear prediction
extrapolates a discrete equally spaced time series $\{y_i\}$ as
\be y_i = \sum_{j=1}^n d_j y_{i-j}\ee
The coefficients $d_j$ are determined by the known data points $\{y_i\}$ by requiring 
that their prediction for
each point $y_i$, based on $y_{i-n} \ldots y_{i-1}$,  vary as little as possible 
from the actual value $y_i$, using
a least-squares criterion. One finds that the $d_j$ are determined from 
correlation functions $\langle y_i y_{i+j}\rangle$,
where the average is over $i$, and the principle computational work in determining $d_j$
is the inverse of a $n\times n$ matrix. In our work we have used $n=20$,
so that the numerical work involved in the extrapolation is 
negligible. 

\begin{figure}[htbp]
\begin{center}
\includegraphics*[width=0.8\hsize,scale=1.0]{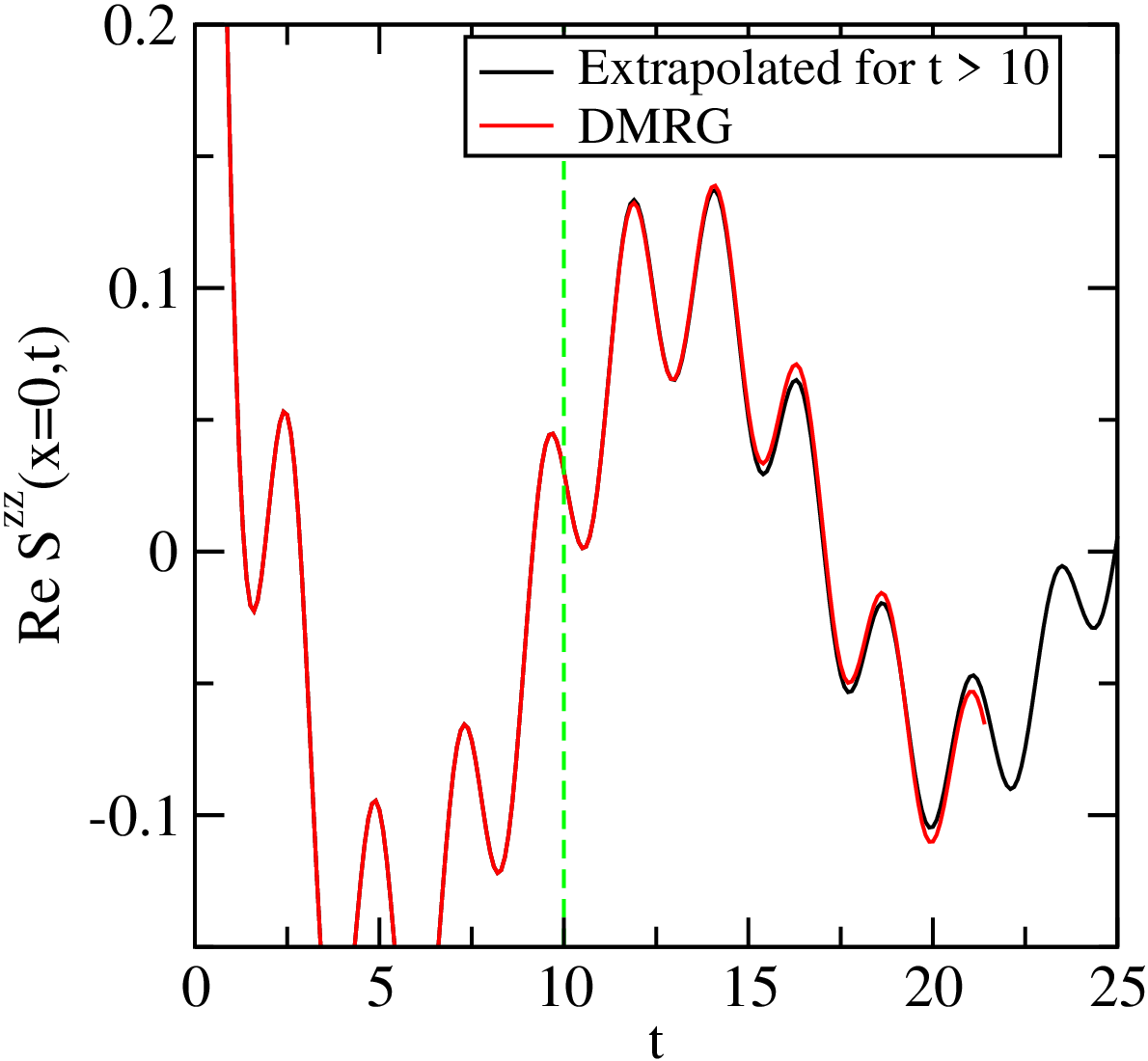}
\caption{DMRG results for the real part of the onsite spin-spin correlation function $S^{zz}(x=0,t)$ for
the more accurate run of Fig. \ref{truncerrors}.
The red curve represents DMRG data out to $t=22$; the black curve is the DMRG data for $t \le 10$ and a linear
extrapolation for $t>10$.}
\label{reSpm}
\end{center}
\end{figure}

In Fig. \ref{reSpm} we show the effectiveness of this extrapolation. For this run which had $\tm=22$,
the data for $t \le 10$ was extrapolated to longer times and compared with the DMRG results. The
error in the extrapolation starts out small and grows reasonable slowly. Provided one does not rely
on too long an extrapolation to try to achieve higher frequency resolution than the data supports,
this method performs much better than ordinary windowing methods.

\begin{figure}[htbp]
\begin{center}
\includegraphics*[width=0.7\hsize,scale=1.0]{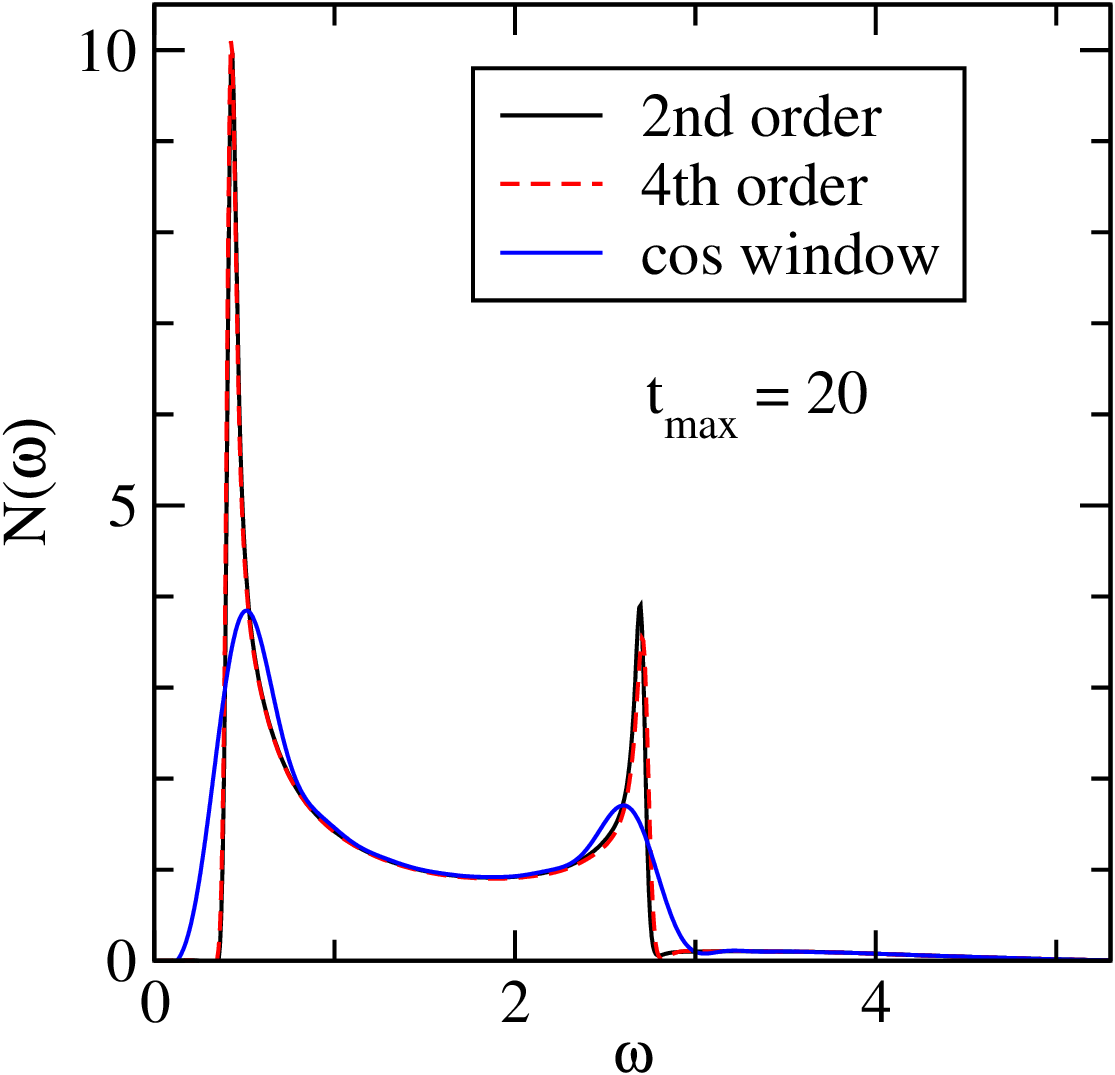}
\caption{Results for the density of states  $N(\omega)$ for the two systems of
Fig. \ref{maxmplot}, with linear extrapolation utilizing only the data out to $\tm=20$. The two curves
are almost identical.
The third curve used the 2nd order data but did not
utilize extrapolation; instead, the data out to $t_{\rm max}$ was multiplied by a simple finite window based 
on the cosine function. }
\label{Nwtau}
\end{center}
\end{figure}

In Fig. \ref{Nwtau} we compare the density of states, defined as $N(\omega) = S^{zz}(x=0,\omega)$,
for the two systems of Fig. \ref{maxmplot}. After the linear extrapolation, the results were
multiplied by a Gaussian window $\exp[-t^2/(20 \tm^2)]$ before Fourier transforming. On this scale
the effects of the Trotter errors in the 2nd order data are almost not visible.  The results are considerably
sharper than with the alternative method which did not use extrapolation. The sharp peaks are
broadened square root singularities from the top and the bottom of the single-magnon dispersion.
Above the top of the band, two- and three-magnon contributions to the spectrum are visible in a small tail.
Using a larger window width
would increase the resolution at the cost of increasing the likelihood of artifacts from the extrapolation.  We
will consistently use this window width. Assuming the extrapolation is accurate, this means that our spectra
should look like the exact spectra, but broadened by convolving with a gaussian
\be
\exp(-\frac{1}{2} \omega^2 / W^2)
\ee
where the frequency resolution is $W = (\tm \sqrt{10})^{-1}$.

Another approach is to fit the moderate time data to
the correct asymptotic long-time form, in this case stemming from the square root singularities,
to extrapolate to long times very accurately.  The linear extrapolation asymptotically gives exponential
decays in time, an incorrect assumption in this case,
so for very long times the fitting method can be more accurate.  A disadvantage of the fitting
method is that it assumes one has some understanding of the results analytically.  Another
disadvantage is that the fitting process can take much more computer time than the linear prediction method,
although still much less than tDMRG simulation itself.
In cases where
one does not know what sort of spectra to expect, one can first fit with the linear prediction method,
and then guess an asymptotic form for fitting.

For the case of $N(\omega)$, we have leading singularities of the form
\be a \theta(\omega-\Delta) (\omega-\Delta)^{-1/2} + b \theta(\Omega-\omega) (\Omega-\omega)^{-1/2}
\label{sing}\ee
where $\Omega$ is the maximum in the single magnon dispersion relation near $\pi/2$ of about 2.725.
This Fourier transforms to long time tails with leading terms of the form
\be A \exp(-i \Delta t) t^{-1/2} + B \exp(-i \Omega t) t^{-1/2}.
\label{singt}\ee
A more convenient form for fitting comes from the Fourier transform
integral identities
\bea \int_{-\infty}^\infty d\omega e^{-i \omega t} &\theta(\omega-b) e^{-a (\omega-b)}
 (\omega-b)^g \nonumber\\=
&\Gamma(1+g)e^{-i b t} (a + i t)^{-1-g} \label{integral}\eea
and
\bea \int_{-\infty}^\infty d\omega e^{-i \omega t} &\theta(b-\omega) e^{-a (b-\omega)}
 (b-\omega)^g\nonumber \\=
&\Gamma(1+g)e^{-i b t} (a - i t)^{-1-g}. \label{integraltwo}\eea
These identities are useful because the frequency expressions have only a single one-sided
singularity,  and convenient well-behaved Fourier
transforms.
Eq. (\ref{integral}) is useful for describing a singularity on a lower edge, while 
Eq. (\ref{integraltwo})
describes an upper edge. Thus to fit to the tDMRG data for the Fourier transform of
$N(\omega)$, we use the asymptotic form
\be A \exp(-i \Delta t) (a+it)^{-1/2} + B \exp(-i \Omega t) (b-it)^{-1/2}.
\label{singtt}\ee
Fitting the accurate 4th order data over the time range $10-20$ with this form, we
find that the 
fit within this range matches the data very accurately, with a typical absolute
deviation of about $2\times 10^{-4}$, or a relative error of about $10^{-3}$.

\begin{figure}[htbp]
\begin{center}
\includegraphics*[width=0.9\hsize,scale=1.0]{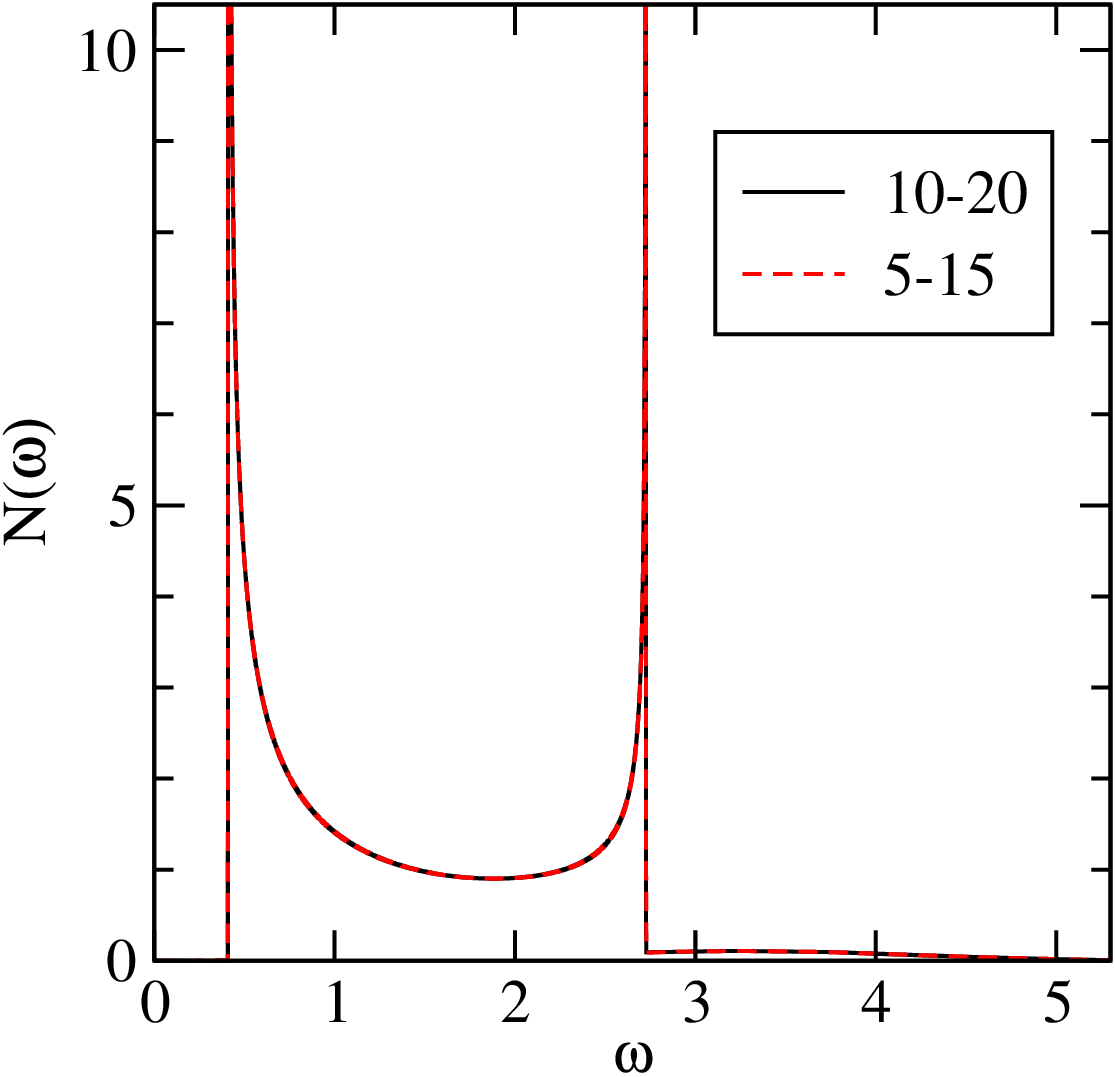}
\caption{Results for the density of states  $N(\omega)$ from fitting to the asymptotic
form Eq. (\ref{singtt}) over the time range indicated in the caption, and then extending
to very large times with the fit before Fourier transforming.  The two curves overlap,
indicating that the error in the determination of $N(\omega)$ is less than the line width. }
\label{Nwfit}
\end{center}
\end{figure}

Using the fitting parameters and asymptotic form, one extends the data to large times.
We did not use the identities Eqs. (\ref{integral}) and (\ref{integraltwo}) to help perform
the Fourier transform (although this might be convenient);
we simply used a fast Fourier transform over a very large range of times (e.g. $-10000 \le t \le 10000$).
Over the fitting region in time a smooth transition is made from the data for 
small times to the fit for large times. The results
of this procedure for $N(\omega)$ are shown in Fig. \ref{Nwfit}. The fits allowed
both $\Delta$ and $\Omega$ to vary; the result for the gap for the time range $10-20$ was
$\Delta = 0.4104327$, accurate to 4 digits (see next section).  The resulting spectra, based on fitting
over different time ranges, appears to be accurate to the line width in the figure.

\section{Single magnon results}

Fourier transforming the DMRG $S(x,t)$ data $x \to k$, we obtain $S(k,t)$ for any $k$. 
The time Fourier transform
gives spectra with (potentially) single magnon and multimagnon contributions. As we discuss
below, for $k_c  < k < \pi$, with $k_c \approx 0.23\pi-0.24 \pi$, a single magnon delta function peak
is present, plus multimagnon continua at higher frequencies. In this section we focus on
this well-defined single-magnon mode.

First consider the band minimum, at $k=\pi$, where the excitation energy is the Haldane gap.
The most accurate method to determine the Haldane gap is still ground state DMRG, where
we have used the same method as in Ref. \onlinecite{huse}, but with up to $m=500$ and $L=400$, to
determine the Haldane gap to very high accuracy, $\Delta = 0.41047925(4)$.  (The
end coupling used to make the lowest excitation have $k=\pm\pi$ was $J_{\rm end}=0.50865$, 
compared to $0.5088$ in Ref. \onlinecite{huse}.)

\begin{figure}[htbp]
\begin{center}
\includegraphics*[width=0.8\hsize,scale=1.0]{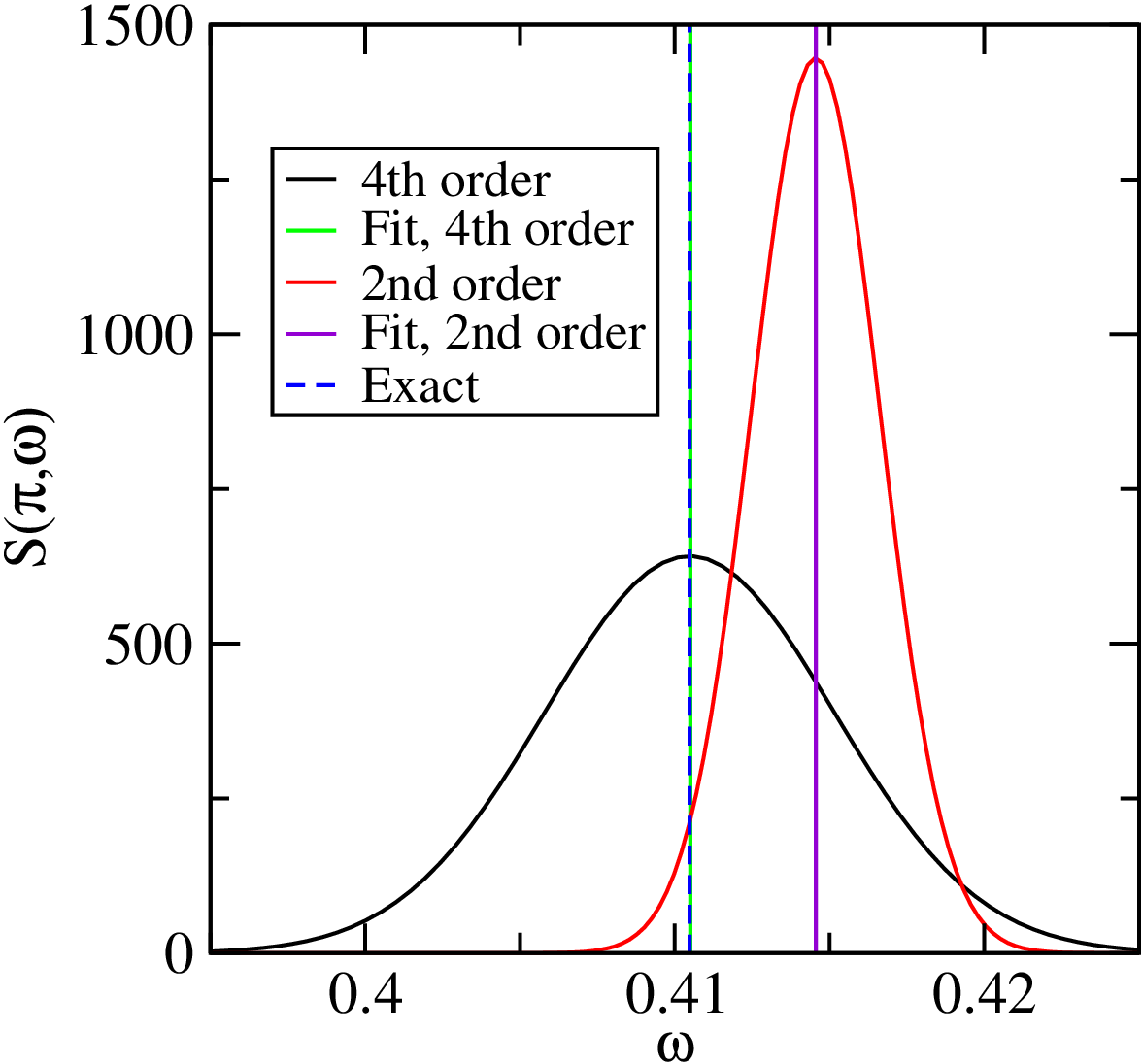}
\caption{Results for   $S(\pi,\omega)$ for the two systems of
Fig. \ref{maxmplot}. The Gaussian shaped curves come from linear extrapolation and Fourier transforming.  The vertical lines represent delta functions coming from a
least squares fit of $A e^{i \omega t}$ to $S(\pi,t)$. The 4th order fit frequency and the ``exact''
ground state DMRG result are indistinguishable in this plot.
}
\label{Spiclose}
\end{center}
\end{figure}

In Fig. \ref{Spiclose}, we show results for $S(\pi,\omega)$ near $\omega=\Delta$.
Linear extrapolation plus Fourier transforming as described above
give rather narrow Gaussian-shaped peaks.
The 2nd order peak is narrower because of a larger $\tm$, and it is shifted from
the exact result because of Trotter error.  Because two and three magnon contributions are very
weak and separated in frequency from $\Delta$, a least squares fit to a pure exponential
is almost identical to the maximum of the broadened peaks. In fact, neglecting any Trotter
shifts, either the maximum or the fit
frequency provide  much more accurate determinations of the exact magnon energy than  the
peak widths would indicate.  For a set of $k$'s spaced $0.01 \pi$ apart we have fit the time data either
to a  pure complex exponential (for $k>1$) and for 
$k_c<k<1$ to a complex exponential
 plus an asymptotic form describing the near-by two-magnon edge. These latter more
 complicated fits are discussed in Section VI.
The  frequency
of the exponential term
determines the dispersion
$\varepsilon(k)$ for $k>k_c$. With this fitting approach, a larger $\tm$ is not very important compared to
the Trotter error, so we utilize the 4th order data. The corresponding result for $\varepsilon(\pi)$
is 0.41050; the error is only a few times $10^{-5}$. For smaller $k$ the error is expected to
be larger because the multimagnon continuum is  larger  relative to the single magnon peak,
and the continuum is closer in frequency,  but the errors for $k>k_c$ are probably no bigger
than $10^{-3}$.

\begin{figure}[htbp]
\begin{center}
\includegraphics*[width=0.8\hsize,scale=1.0]{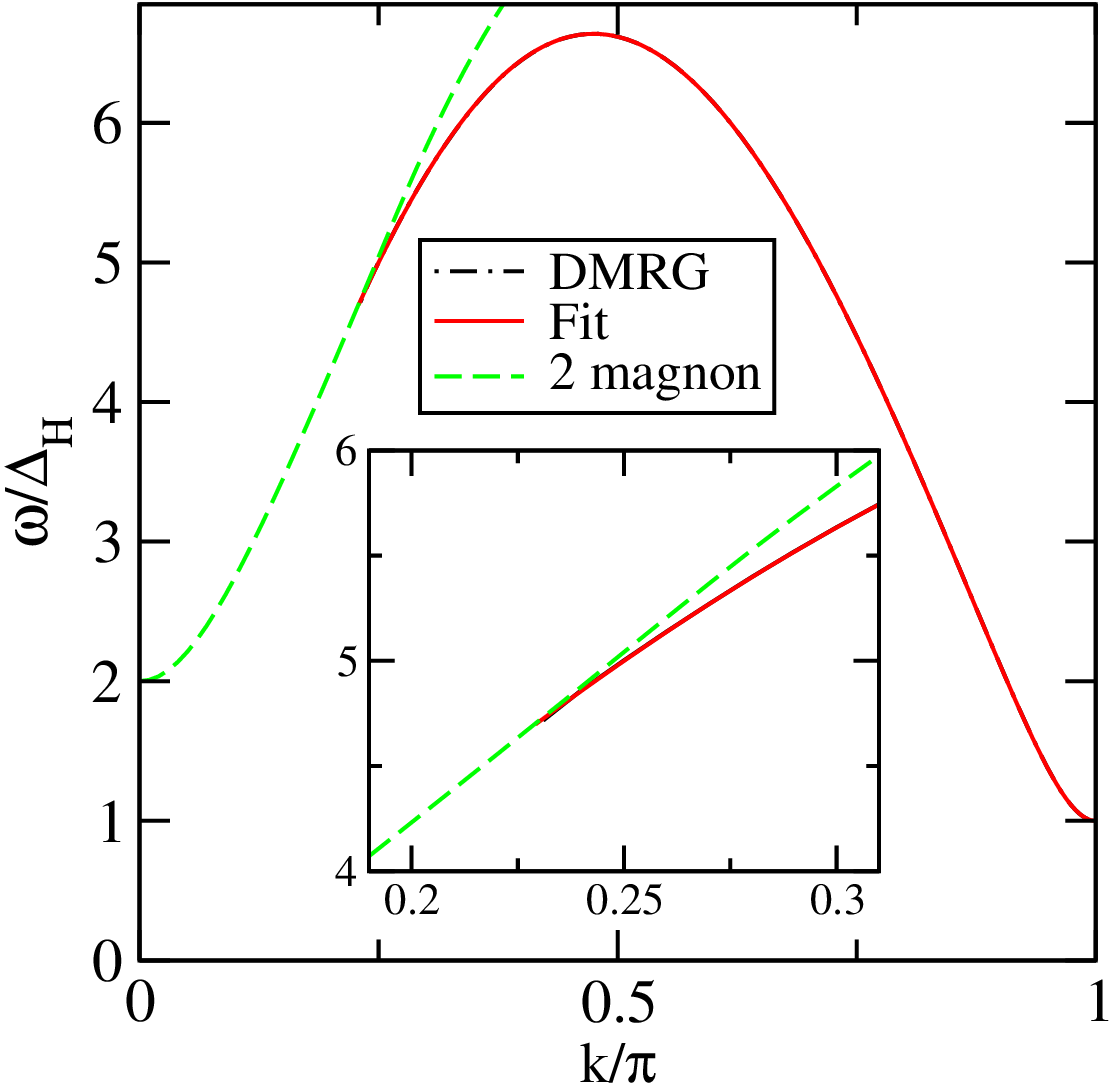}
\caption{Single magnon dispersion. The curve labeled ``DMRG''  comes from the fit of a pure
exponential to the DMRG $S(k,t)$ data. The curve labeled ``Fit'' is the analytic expression
Eq. (\ref{fitdispersion}). 
The final curve is the two magnon band minimum at $2 \varepsilon(\pi-k/2)$.
The inset shows the region near $k_c$, where the magnon line enters the two magnon continuum.
The first two curves are not very meaningful well below $k_c$, since the there is no single 
magnon delta function. }
\label{dispersion}
\end{center}
\end{figure}

\begin{figure}[htbp]
\begin{center}
\includegraphics*[width=0.6\hsize,scale=1.0]{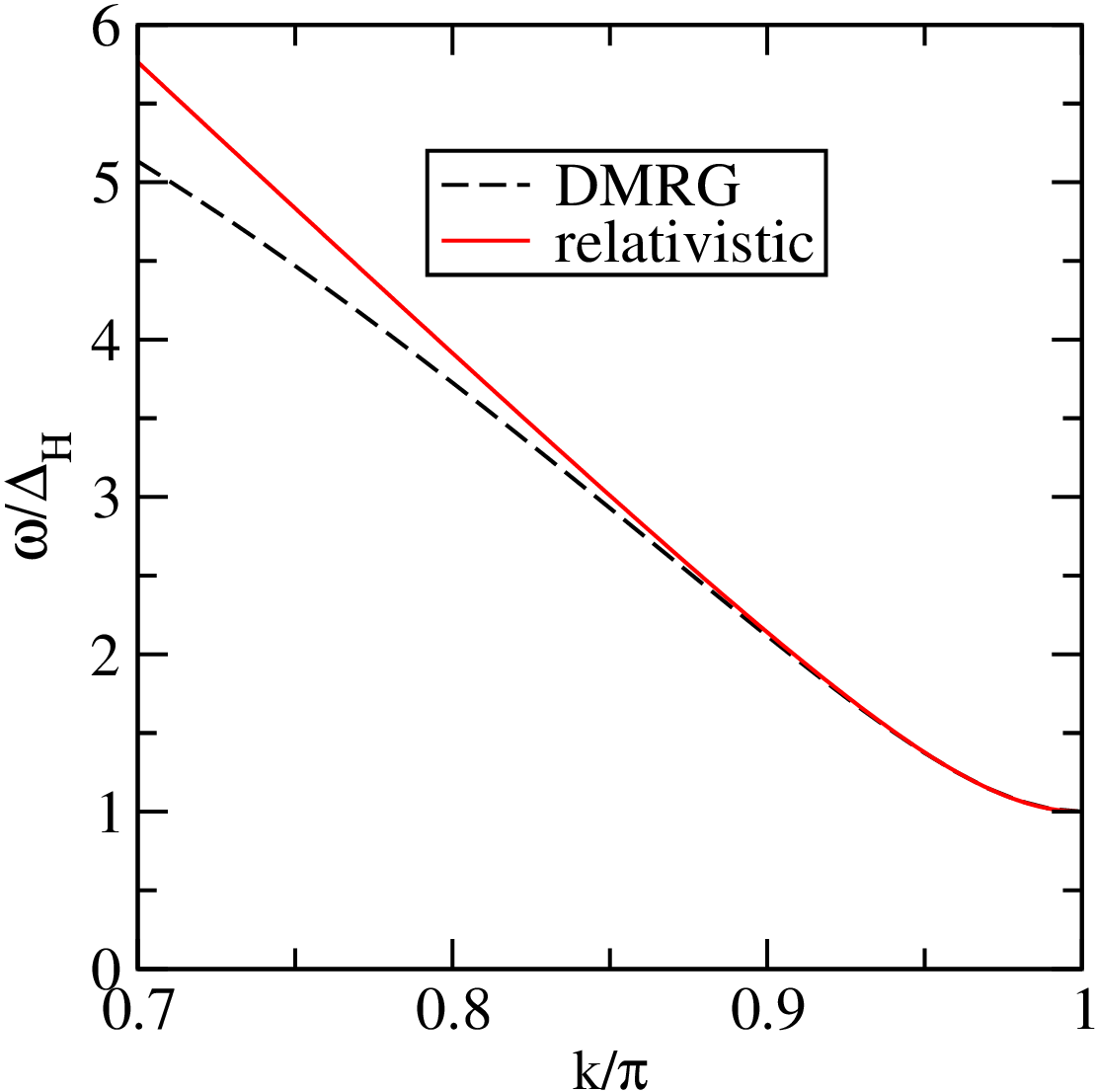}
\caption{Comparison of the single magnon dispersion from DMRG and the relativistic approximation,
Eq. \ref{Ek}. }
\label{disp_near_pi}
\end{center}
\end{figure}

In Fig. \ref{dispersion} we show $\varepsilon(k)$ from the 4th order run, along with an
analytic fit motivated by the NL$\sigma$ model.
The massive triplet excitations of the NL$\sigma$M have the 
relativistic dispersion relation:
\be \epsilon_0(\tilde k)=\sqrt{\Delta^2+v^2\tilde k^2}.\label{Ek}\ee
Here the momentum $\tilde k$ of the NL$\sigma$M is $k-\pi$ for the spin chain.  
In the NL$\sigma$M, $\tilde k$ can take any real value.  Of course, 
in the spin chain, crystal momenta lie in the Brillouin zone, $|k|<\pi$. 
This discrepancy limits the validity of the NL$\sigma$M, 
especially when we consider multi-particle excitations. 
$\epsilon_0(\tilde k)$ is only an approximation to the exact single magnon 
dispersion relation, $\varepsilon(\tilde k)$.
 We expect a perfectly stable single 
magnon excitation to exist for $\tilde k<\pi -k_c$ with this dispersion relation. 
Strictly speaking $\varepsilon(\tilde k)$ is not defined for $\tilde k>\pi -k_c$.  
The ``Fit'' curve shown in Fig. \ref{dispersion} is based on the 
expression
\be \epsilon (k) \approx 
 \Delta\sqrt{1+\sum_{n=1}^5a_n\{ 1-\cos [n(\pi -k)]\}}.\label{fitdispersion}\ee
with 
the parameters $a_n$ given in Table 1.  The gap, at $k=\pi$,  from the data used in the fit  
is $\Delta =0.410504$; one could also use the more accurate value $\Delta =0.41047925$.  
$\epsilon (k)$ goes through a maximum of $2.72551$ at $k\approx .476\pi$ 
and has the value $1.96$ at $k\approx 0.23 \pi$ near where the single magnon excitation becomes 
unstable. There is an inflection point ($d^2\epsilon /dk^2=0$) at $k_{\rm in}\approx 0.868 \pi$. 
As shown in Fig. \ref{disp_near_pi} 
$\epsilon (\tilde k)$, agrees quite well with the Lorentz invariant 
approximation, $\epsilon_0(\tilde k )$, for $\tilde k <0.1\pi$ , 
and reasonably well for $\tilde k <0.2\pi$,
with $v\approx 2.472$. 
Also shown in Fig. \ref{dispersion} is the two magnon band minimum, which for 
$|k| < 2(\pi-k_{\rm in}) \approx 0.265 \pi$ is given
by $2 \varepsilon(\pi-k/2)$. Near 
$\tilde k\approx 0$, this is approximately
\be \epsilon (\tilde k)\to \Delta + {(v\tilde k)^2\over 2\Delta }.\ee
The intersection of the two magnon minimum and the single magnon dispersion line
determines $k_c$, as show in the inset of Fig. \ref{dispersion}.

\begin{figure}[htbp]
\begin{center}
\includegraphics*[width=0.6\hsize,scale=1.0]{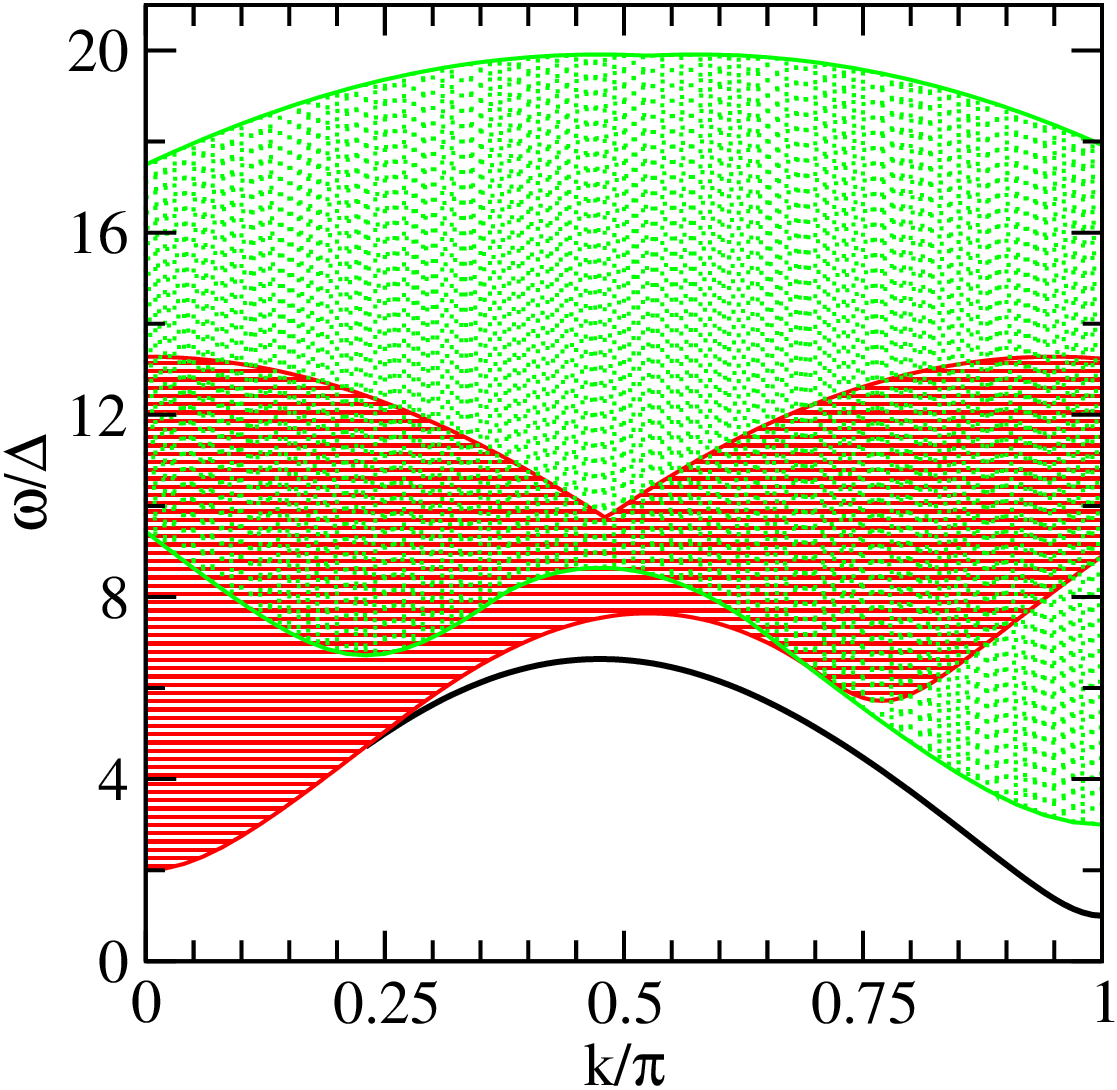}
\caption{Single magnon line (black solid line), two magnon band (horizontal line fill, red),
and three magnon (dotted fill, green) bands. }
\label{magnon_minmax}
\end{center}
\end{figure}

The single magnon dispersion can be used to construct the multimagnon band minima and maxima,
assuming that the magnon-magnon interactions are negligible, and adding the single magnon energies. 
The resulting bands are shown in
Fig. \ref{magnon_minmax}. In this construction, it was assumed that the single magnon line stops abruptly
at $0.23\pi \approx k_c$. This resulted in slope discontinuities visible in the two magnon band. However,
at $k_c$, as we discuss below, the sharp "single magnon peak"
can be described as either a single magnon or two magnon feature.  This ambiguity effectively blurs the
distinction between the two and three magnon bands near the slope discontinuities. If one changed the construction
to include the broadened ``single magnon peak" to values of $k$ below $k_c$, the two magnon bandwidth would
be broadened near the slope discontinuities. This suggests that, for example, at $k=\pi$ one might expect to
see the two magnon band maximum as a visible feature of the spectrum, but one might not see anything
for the two magnon minimum. (The spectrum at $k=\pi$ is shown in Section V.)

\begin{table}[b]
\caption{Coefficients in the fit of the single magnon dispersion relation, Eq. (\ref{fitdispersion}). \label{tableI}}
\begin{tabular}{cccccccc}
\hline\hline 
$n$&
$a_n$\\
1 & 1.96615 \\
2 & 21.20162 \\
3 & -1.61279 \\
4 & -0.04766\\
5 & 0.02407\\
\hline 
\hline\hline
\end{tabular}
\end{table}

\begin{figure}[htbp]
\begin{center}
\includegraphics*[width=0.6\hsize,scale=1.0]{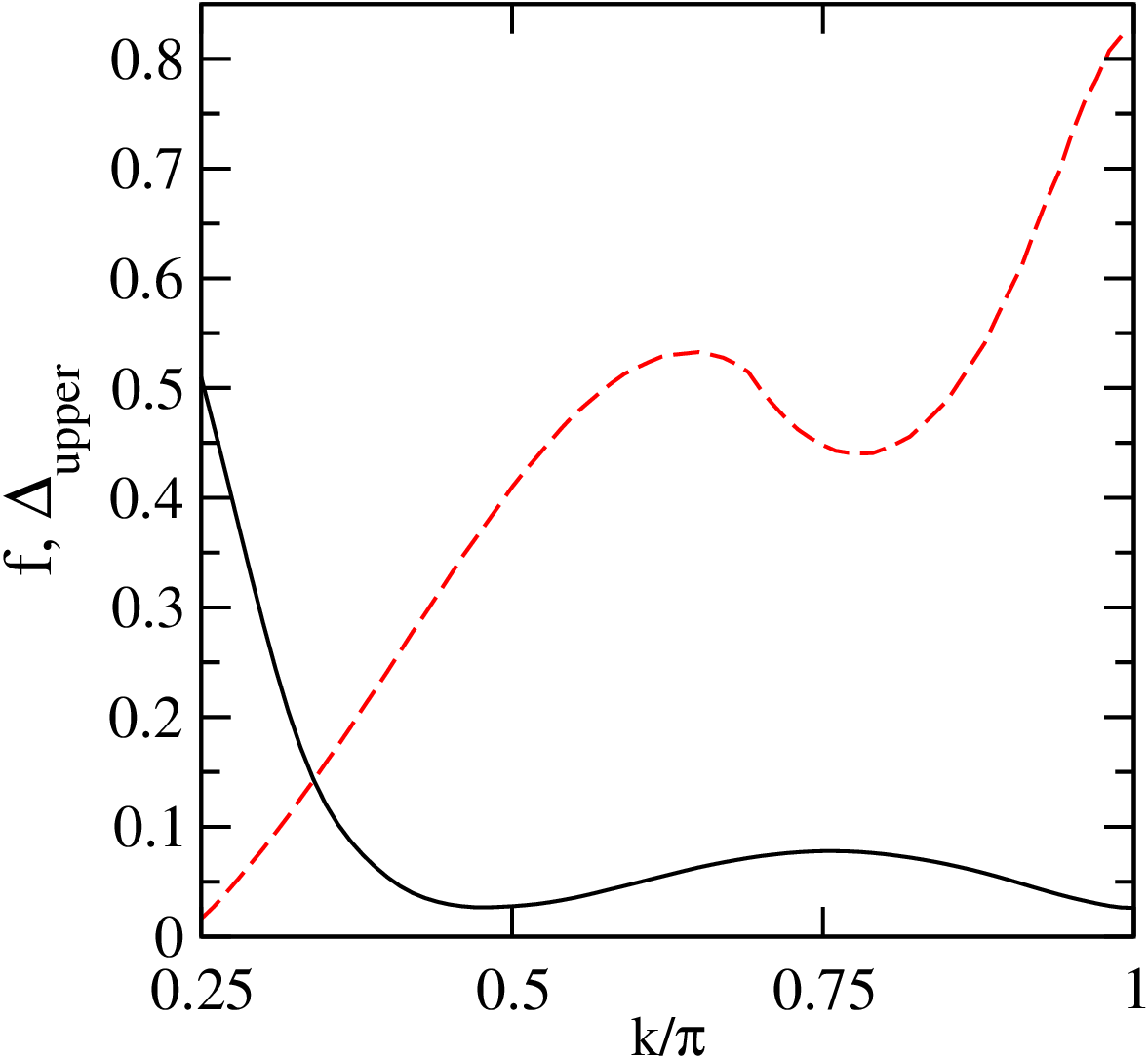}
\caption{Fraction of the spectral weight in the multimagnon continua for $k>k_c$,
shown by the solid black line.  The gap between the single magnon line and the
lowest multimagnon band is shown by the dashed red line.}
\label{fratio}
\end{center}
\end{figure}

The amplitude of the single magnon peak is approximated well by\cite{weston}
\be S(q,t=0) \approx \frac{v Z }{\sqrt{v^2 k^2 + \Delta^2}}
\ee
with $Z=1.26$. A related  quantity is the fraction of the spectral weight in the multimagnon
continua, given by 
\be f = \frac{S(q,t=0)-A}{S(q,t=0)} \ee
where $A$ is the amplitude of the single magnon peak. This quantity is shown in Fig. \ref{fratio}.
It is interesting that $f$ shows non monotonic behavior with $k$, with  minima near
$k=\pi/2$ and $k=\pi$. This non monotonic behavior is roughly correlated with the gap between
the single magnon line and the lowest multi-magnon band minimum.  Naively, one might
expect that a nearby continuum has an easier time than a faraway one does in
taking spectral weight from the single magnon line.

\section{$k\approx 0$}
We now discuss the properties of the spectral function for small $k$. That
$S\propto k^2$ as $k \to 0$
can easily be proven to be exactly true using the fact that 
$\sum_jS^z_j|0\ket =0$ (singlet ground state) and Taylor expanding 
$S(k,\omega )$ in Eq. (\ref{S}) to second order in $k$. 
On general symmetry grounds, based on the NL$\sigma$M, we expect that 
near $k\approx 0$, S will contain only multi-particle continua 
corresponding to  even numbers of bosons.  At a finite small $k$, 
the lowest energy 2-magnon state 
with total momentum $k$ is one in which each magnon has momentum $\pi+k/2$.  
Therefore, since the dispersion is even about $\pi$,
 the bottom of the 2-magnon continuum should be 
exactly at $2\epsilon (\pi-k/2)$. This remains true up to 
$k=2(\pi -k_{\rm in})\approx .265\pi$ (recall $k_{\rm in}\approx 0.868\pi$ is the inflection point). 
For a range of energies, at small enough $k$, the {\it only} 
possible excitations have 2 magnons.  This is true 
up to $\omega = 4\epsilon (k/4)$, for $k<(\pi -k_{\rm in})$.

A simplified, ``mean field'' version of the NL$\sigma$M is 
a free massive boson model with Lagrangian:
\be \mathcal{L}={1\over 2v}[(\partial_t\vec \phi )^2-v^2(\partial_x\vec \phi )^2-\Delta^2(\vec \phi )^2],
\label{Lfree}\ee
and no constraint on $\vec \phi$. 
Expanding $\vec \phi$ in boson creation and annihilation operators one finds the 
free boson result
\be S_0(k,\omega )\approx 
{k^2\sqrt{\omega^2-(kv)^2-4\Delta^2}\over v[\omega^2-(kv)^2]^{3/2}}\theta (\omega^2-(vk)^2-4\Delta^2).
\label{S0}\ee
As expected from general principles, this vanishes quadratically as $k\to 0$, 
and also vanishes below the 2-magnon threshold, $\omega_{\rm th} = 2\sqrt{\Delta^2+(vk/2)^2}$. 
The exact 2-magnon expression for $S(k,\omega )$, in the NL$\sigma$M is 
known exactly and can be written:
\be S_{0\sigma}(k,\omega )=S_0(k,\omega ){\pi^4\over 64}{1+(\theta /\pi )^2\over 1+(\theta /2\pi )^2}
\left({\tanh \theta /2\over \theta /2}\right)^2,\label{Ssig}\ee
where the rapidity, $\theta$, is defined by:
\be \theta = 2\cosh^{-1}[(\omega^2-v^2k^2)/(4\Delta^2)].\ee

\begin{figure}[htbp]
\begin{center}
\includegraphics*[width=0.7\hsize,scale=1.0]{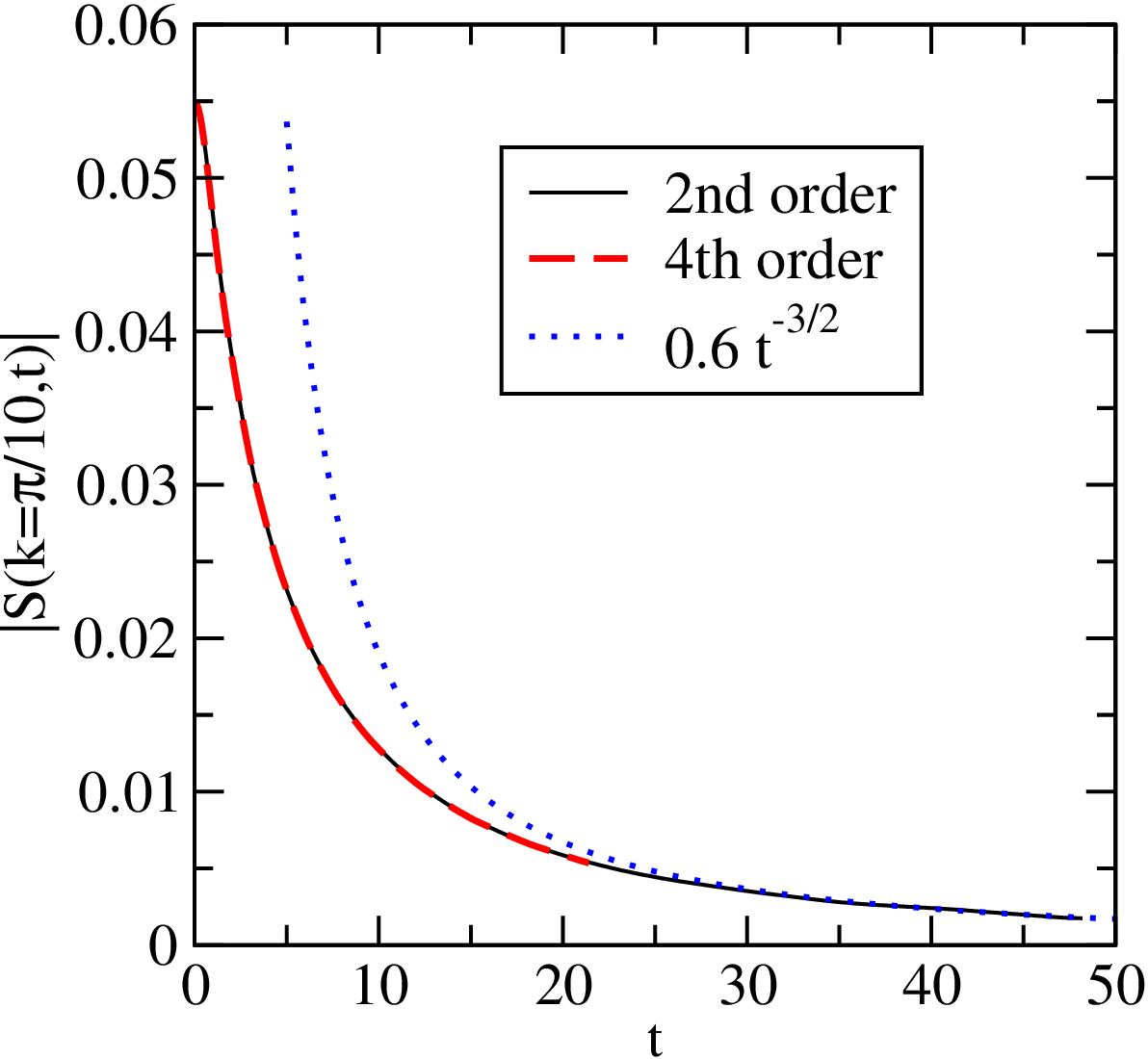}
\caption{Comparison of  $|S(k,t)|$ for $k=\pi/10$ for two different runs, with 2nd order and
4th order Trotter decomposition runs of Fig. 2, and the asymptotic form expected analytically.}
\label{Stentht}
\end{center}
\end{figure}

\begin{figure}[htbp]
\begin{center}
\includegraphics*[width=0.7\hsize,scale=1.0]{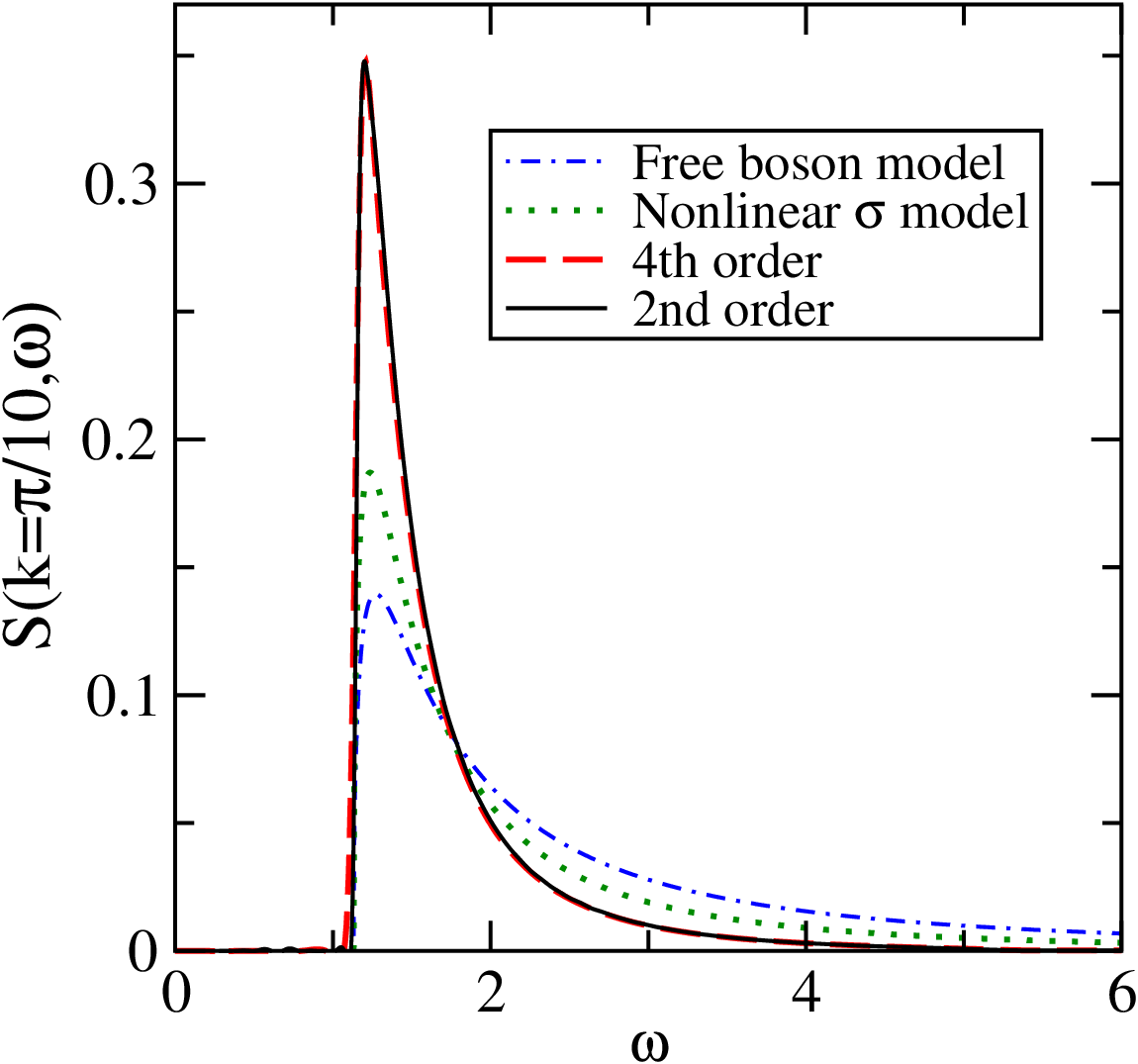}
\caption{$S(k,\omega)$ for $k=\pi/10$ for two different runs, with different accuracies and different
total times, both Fourier transformed using linear extrapolation.  For comparison, two analytic
results, based on Eqs. (\ref{S0}) and (\ref{Ssig}), are shown.}
\label{ktenthcomp}
\end{center}
\end{figure}

Just above $\omega_{\rm th}$, both
Eqs. (\ref{S0}) and (\ref{Ssig}) rise as $(\omega-\omega_{\rm th})^{1/2}$, which would lead to
the asymptotic time behavior 
\be S(k,t) \sim e^{-i \omega_{\rm th} t} t^{-3/2}.\ee
In Fig. \ref{Stentht} we compare $|S(k,t)|$ for a typical small value of $k$
with  $A t^{-3/2}$, finding good agreement for large $t$
with the empirical parameter $A=0.6$.
Using the linear prediction method, we Fourier transformed the $S(k,t)$ DMRG results for both
runs, and in  Fig. \ref{ktenthcomp}, they are compared to Eqs. (\ref{S0}) and (\ref{Ssig}).
They look qualitatively similar. In particular, 
the threshold singularity appears the same and 
they have peaks at similar frequencies. 
However, the peak is about a factor of 2 larger in the DMRG 
data than in the NL$\sigma$M. 
 Furthermore, the field 
theory results drop off much more slowly at large $\omega$. This 
latter feature is to be expected since magnons of arbitrarily 
high momentum are included in the field theory while 
there is a cut off at the Brillouin zone boundary in reality. 

\begin{figure}[htbp]
\begin{center}
\includegraphics*[width=0.7\hsize,scale=1.0]{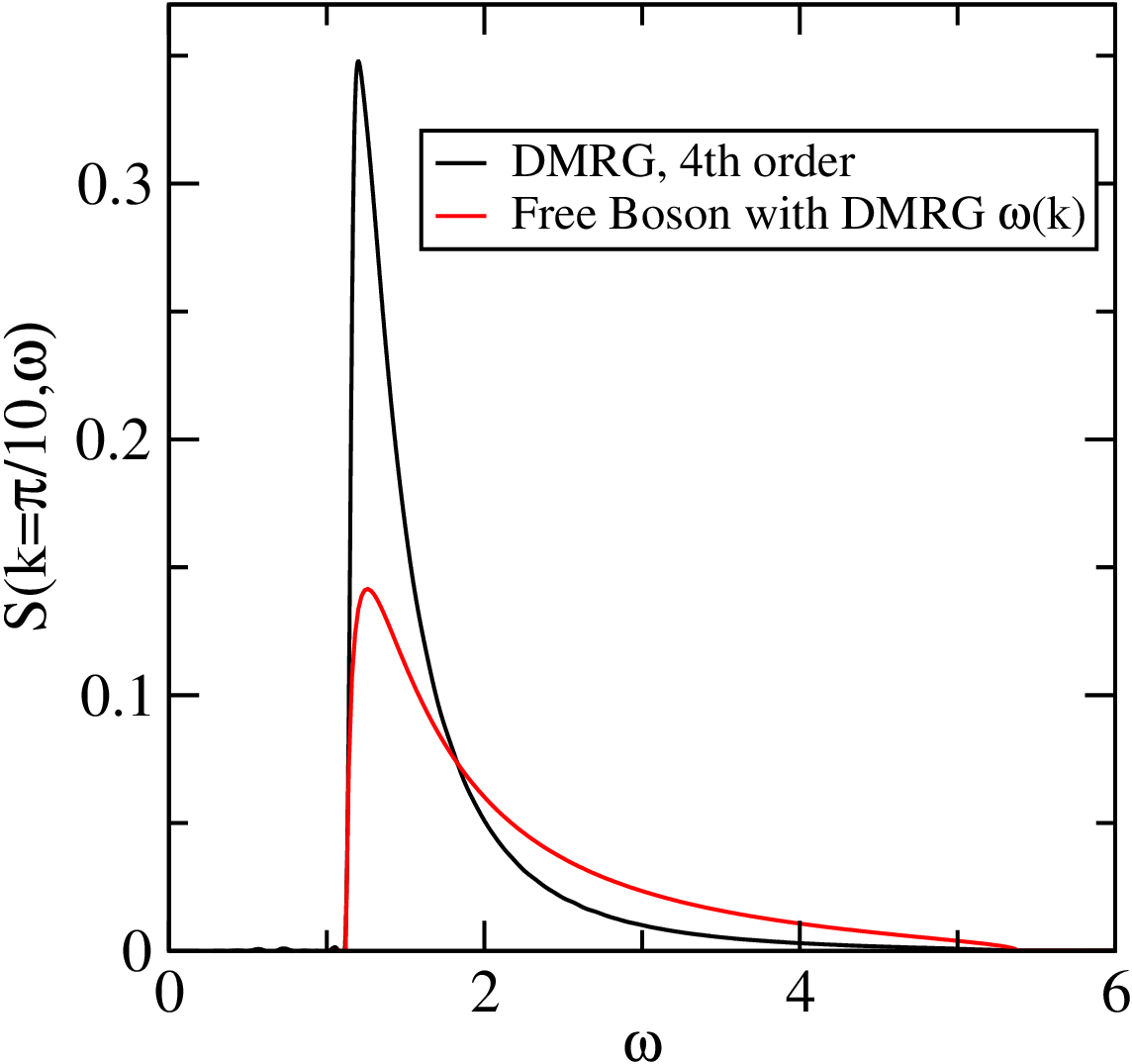}
\caption{$S(k,\omega)$ for $k=\pi/10$, comparing the near-exact DMRG results with the modified
free boson result.}
\label{ktenthcompw}
\end{center}
\end{figure}

A simple way to improve the high frequency behavior 
in the free boson approximation is to replace the relativistic 
model of Eq. (\ref{Lfree}), by a Hamiltonian:
\be H={1\over 2}\sum_k[\vec \Pi_k\cdot \vec \Pi_{-k}+
\epsilon^2(k)\vec \phi_k\cdot \vec \phi_{-k}],\ee
where $\Pi^a_k$ is canonically conjugate to $\phi^a_k$ 
and $\epsilon (k)$ is the numerically determined single 
magnon dispersion relation. The appropriate form 
of the small $k$ spin operators:
\be \vec S_k\approx \sum_{k'}\vec \phi_{k'}\times \vec \Pi_{k-k'},\ee
is determined by the requirement that the spin commutation relations 
are obeyed and that $\vec S_0$ commute with the Hamiltonian. This 
changes $S_0(k,\omega )$ to:
\be S_0\to {[\epsilon (k')-\epsilon (k-k')]^2\over 
2\epsilon (k')\epsilon (k-k')|\epsilon '(k')-\epsilon '(k-k')|},\label{S0re}\ee
where $\epsilon (k)$ is the exact (numerically determined) 
dispersion relation and $\epsilon '(k)$ denotes its derivative. $k'$, 
the momentum of one of the 2 magnons, 
is determined from $\omega$ and $k$ by energy-momentum conservation:
\be \epsilon (k')+\epsilon (k-k')=\omega .\label{k'}\ee
For low enough energy, there is only one pair of solutions to Eq. (\ref{k'}), 
and one element of the pair should be chosen in evaluating Eq. (\ref{S0re}). 
The resulting improvement of the free boson result is shown in Fig. \ref{ktenthcompw}
As expected, there is little change near the peak and threshold, 
but the high energy tail is cut off. (Here we restricted 
each boson to have $|\tilde k|<\pi /2$.)

\section{$k\approx \pi$}
\begin{figure}[htbp]
\begin{center}
\includegraphics*[width=0.7\hsize,scale=1.0]{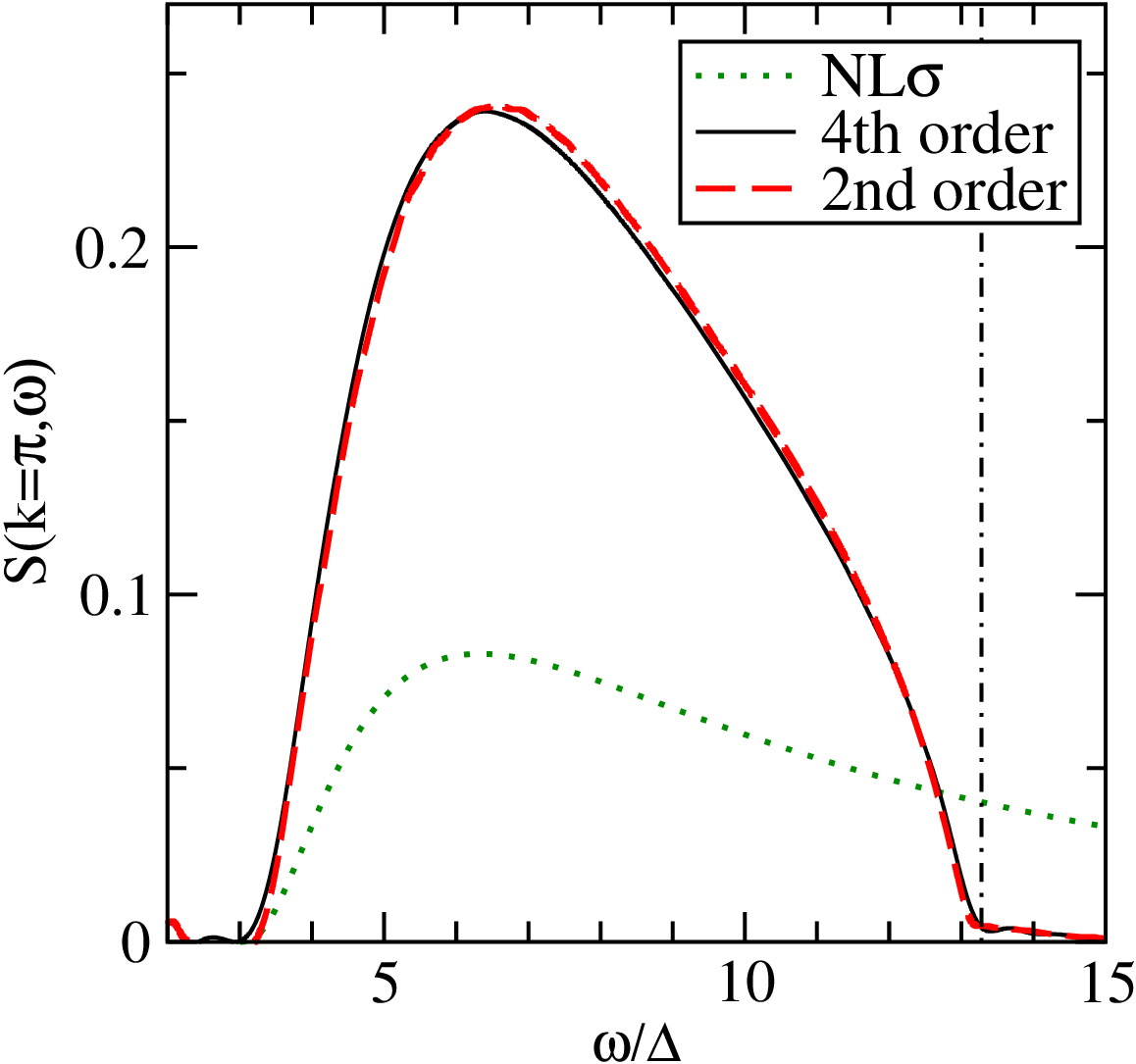}
\caption{$S(k,\omega)$ for $k=\pi$ in the multimagnon frequency regime
for two different runs. Also shown are the results from the  NL$\sigma$M model.
The vertical line near $\omega/\Delta \approx 13$ is two times the maximum of the single magnon
dispersion, roughly locating the top of the two magnon band.
The tiny bump near $\omega/\Delta=2$ is an artifact of the FT of the much larger
single magnon peak at
$\omega/\Delta=1$, while we believe the small tail above $\omega/\Delta \approx 13$ is real.}
\label{kpi3mag}
\end{center}
\end{figure}

In addition to the single magnon mode, which has most of the spectral 
weight near $k=\pi$ there is also a contribution from 
$3$, $5$, $\ldots$ magnons. Of these, the largest is expected 
to be the $3$-magnon contribution. The lower threshold 
for this, corresponding to each magnon having momentum $\tilde k/3$ 
is at $3\epsilon (\tilde k/3)\approx 3\sqrt{\Delta^2+(v\tilde k/3)^2}$. 
At small $\tilde k$ the exact lower threshold approaches $3\Delta + (v\tilde k)^2/(6\Delta )$. 
Note that the presence of a multi-magnon contribution is a consequence 
of inter-magnon interactions; it vanishes for the non-interacting model 
of Eq. (\ref{Lfree}).
The exact 3-magnon form factor is known for the NL$\sigma$M and the resulting 
3-magnon contribution to $S$ can be expressed 
in terms of an elementary integral. The result is compared to our 
DMRG results in Fig. \ref{kpi3mag}.  Again there is a qualitatively 
similarity, with a peak at a similar energy, but now the NL$\sigma$M 
peak is about $3$ times too low and there is far too much 
spectral weight at high energies. The total  spectral 
weight in the 3-magnon peak compared to that 
in the single magnon is found from DMRG to be 2.7\%.
The lower edge of the multimagnon band is given by the three magnon edge, c.f. Fig. \ref{magnon_minmax}.
The band has a sharp dropoff at the two magnon band maximum.  The three magnon band above that
is rather small. As discussed earlier, one does not expect a sharp feature for the two magnon minimum
(nominally near $\omega/\Delta \approx 9$), and none is visible.

\section{$k\approx k_c$}
\begin{figure}[htbp]
\begin{center}
\includegraphics*[width=0.7\hsize,scale=1.0]{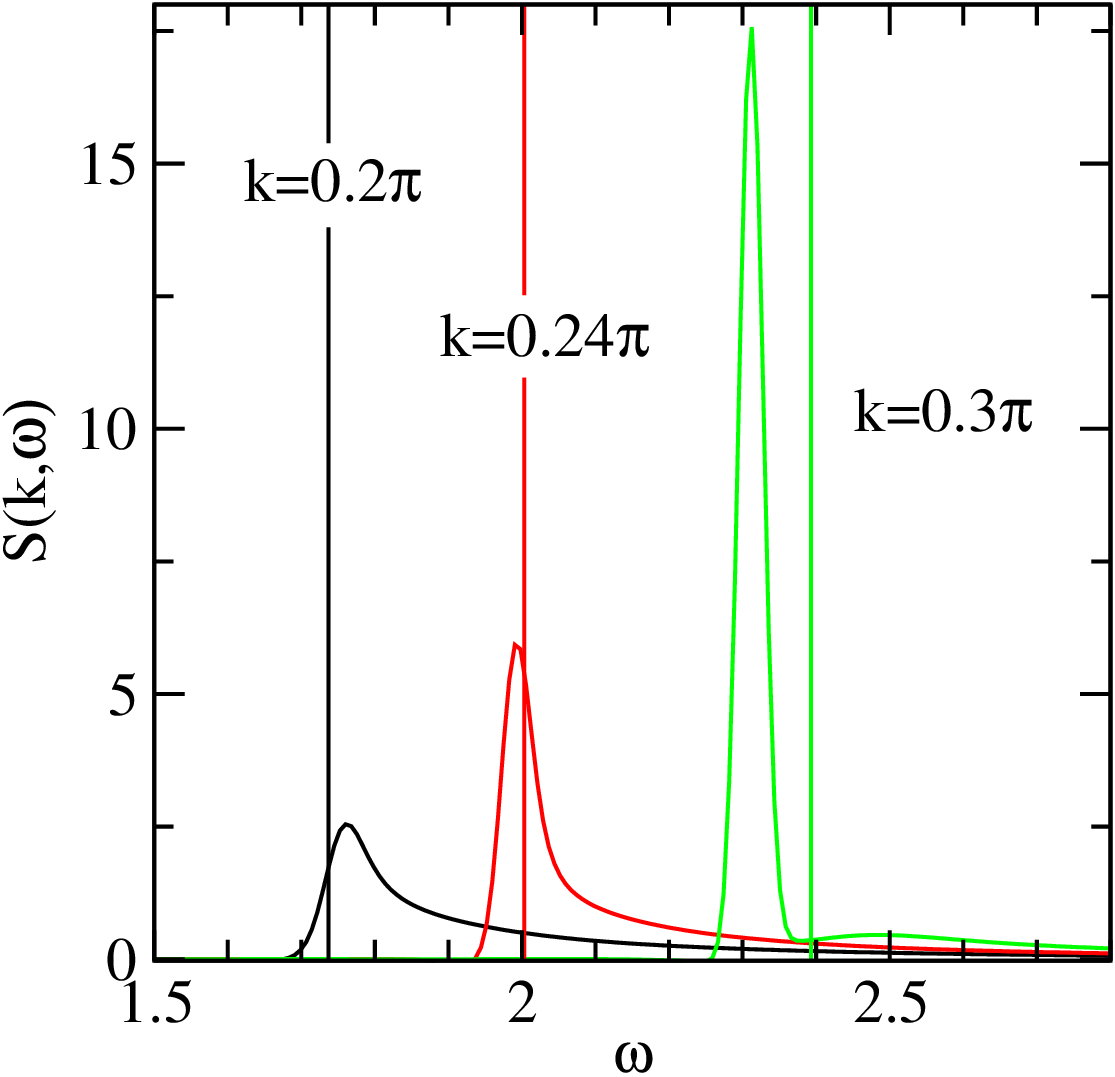}
\caption{$S(k,\omega)$ near $k_c \approx 0.23\pi-0.24 \pi$ for the 4th order run using the linear prediction method. 
The vertical line associated with each $k$ is  the two magnon lower band edge. }
\label{Snearkc}
\end{center}
\end{figure}

A remarkable feature of $S(k,\omega )$ which is completely 
missed by the NL$\sigma$M approach is the merging of the 
single particle peak into the 2-particle continuum at $k=k_c\approx 0.23\pi-0.24\pi$. 
Results for $S(k,\omega)$ from
the linear prediction method near
$k_c$ are shown in Fig. \ref{Snearkc}. Above $k_c$, one sees the separate single magnon peak, broadened by
the finite run time and Fourier transform. By $k=0.2\pi$ the peak has disappeared, and one sees a characteristic
small $k$ line shape. Close to $k_c$ the broadening from the finite maximum time obscures the details of
the spectrum.

\begin{figure}[htbp]
\begin{center}
\includegraphics*[width=0.7\hsize,scale=1.0]{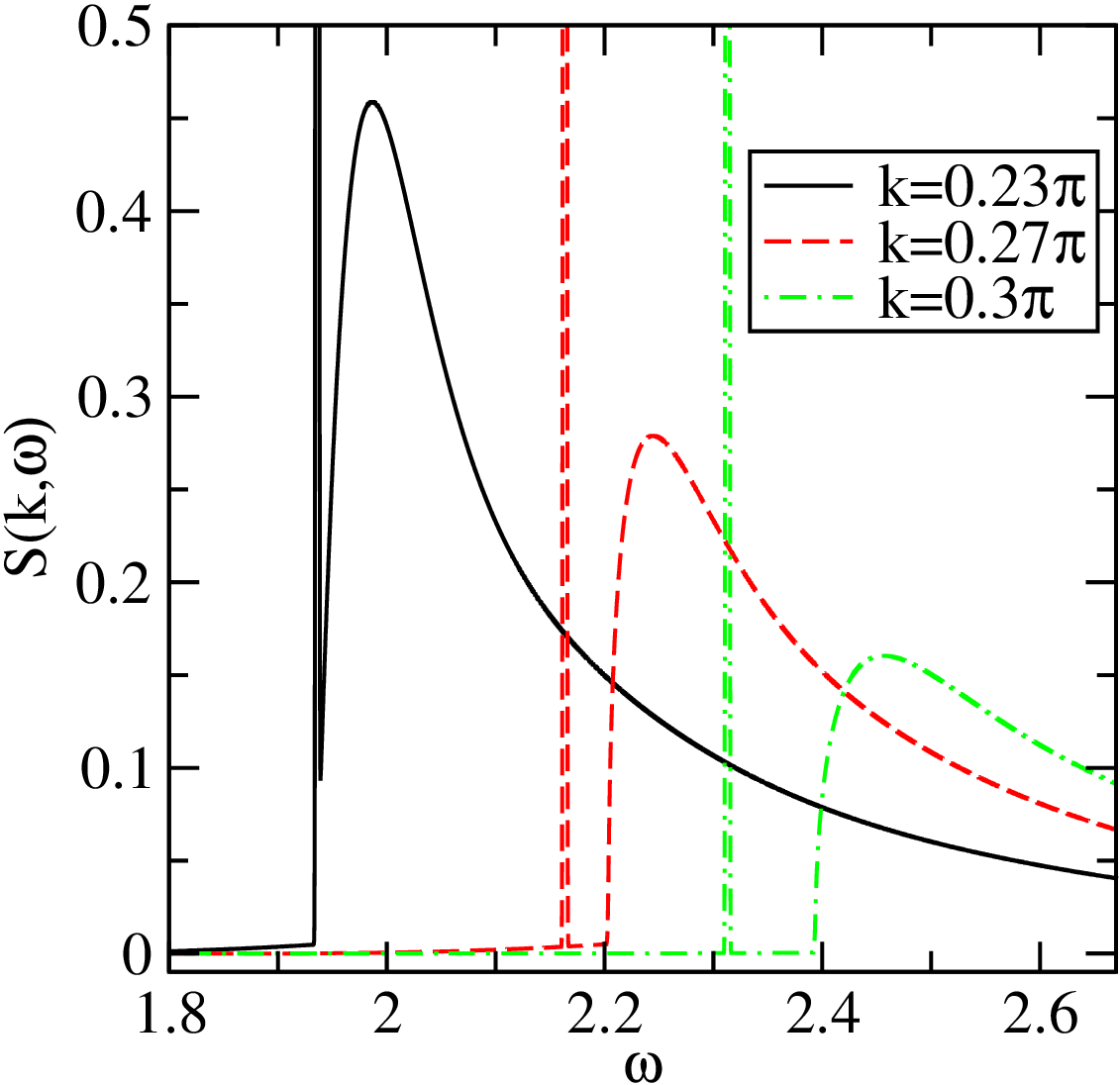}
\caption{$S(k,\omega)$ near $k_c \approx 0.23\pi-0.24 \pi$ for the 4th order run using fitting to the asymptotic
time decay. }
\label{Snearkcfit}
\end{center}
\end{figure}

The fitting method is very useful to capture the behavior of $S(k,\omega)$ near $k_c$ more accurately.
Just above $k_c$ we expect a combination of a single magnon $\delta$ function peak and a two-magnon
continuum similar to that at small $k$. To fit the time data as accurately as possible, we assume
that the two magnon band starts exactly at the expected threshold 
$\omega_{\rm th}= \min_{k'} \epsilon(\pi-k/2+k') + \epsilon(\pi-k/2-k')$. The relevant values of $\epsilon(k)$
used to determine $\omega_{\rm th}$ are far from $k_c$ and  a simple exponential fit determines the
magnon peak location very accurately. We also assume that the small $k$ threshold behavior 
$(\omega-\omega_{\rm th})^{1/2}$ applies.  Including, in addition, the next expansion term 
$(\omega-\omega_{\rm th})^{3/2}$, we utilize the fitting form (cf Eq. (\ref{singtt}))
\bea 
&A \exp(-i \bar \omega t) +
B \exp(-i \omega_{\rm th} t) (b+it)^{-3/2} \nonumber\\&+ C \exp(-i \omega_{\rm th} t) (c+it)^{-5/2}.
\label{fitkc}\eea
Results from this fitting followed by Fourier transforming are shown in  Fig. \ref{Snearkcfit}. The
fits deviated from the data over the range $t=10-20$ typically by a few times $10^{-5}$, and the
magnitude of the data points fitted to was typically near $0.1$---an excellent fit, making a convincing case
that the assumed asymptotic form is correct and that the results for $S(k,\omega)$ are very accurate.
The peak locations from these fittings were used in the determination of the dispersion relation of
Section III near $k_c$.

One can view 
the split-off of the single magnon peak from the two particle continuum at $k_c$ in two different ways.
First, in an approach motivated by the small $k$ two-magnon nature of the spectrum, one can
regard the splitoff  as due to the formation 
of a sharp 2-magnon bound state for $k>k_c$.\cite{huse}
Second, if we imagine that $k_c$ is ``large'' so that the two magnon description
is inappropriate, then we can regard the single magnon peak as surviving for $k$ near but less than $k_c$, but with
a broadening caused by decays into two magnons.  Here we will consider in more detail these two pictures.

Consider first the two-magnon bound state picture.
Since the quantum numbers of the single magnon state with $S_z=1$
and a two magnon bound
state composed of a $S_z=0$ and a $S_z=1$ magnon are identical, we are free to regard the excitation which splits
off in either way.  Well above $k_c$ the two-magnon bound state picture is clearly 
not very useful, since its formation would imply large
magnon-magnon interactions which are not otherwise observed.
The NL$\sigma$M 
does not have any bound states.  Furthermore, 
in {\it any} Lorentz invariant theory, increasing the centre of mass 
momentum can never lead to bound state formation. As was 
discussed in Ref. \onlinecite{huse}, since the bosonic magnon excitations 
form a triplet, and any excitation produced from the singlet ground state by 
the spin operators must also be a triplet, it follows that 
the bound state wave-function must be antisymmetric in its 
spatial coordinates. In  Ref. \onlinecite{huse}, it was found that the
effective magnon-magnon interaction is {\it attractive} 
in the antisymmetric channel. On the other hand, it is 
apparently repulsive in the symmetric channel. This is 
related to the BEC picture of the transition at a 
critical magnetic field\cite{sorensen} where the Zeeman energy equals $\Delta$. 

Even in one dimension, an arbitrarily weak attraction does 
not produce a bound state in the antisymmetric channel (although 
it does produce one in the symmetric channel.) It was 
observed in Ref. \onlinecite{huse} that as $k$ is increased from zero, 
the momenta of the two magnons forming the bound state near 
the threshold $\omega_{min}(k)$, which are near 
$\pi + k/2$, approach the inflection points $k_{in}\approx 1.131\pi$. 
To study the bound state near the threshold, we 
can expand the dispersion relation, $\epsilon $, near 
$\pi +k/2$.  The momenta of the two bosons are:
$\pi +k/2\pm q$.  Expanding the total kinetic 
energy in powers of $q$ gives
 the effective 
kinetic energy :
\be T\approx 2\epsilon +
\epsilon^{(2)}q^2
+{\epsilon^{(4)}\over 12}q^4+\ldots ,\ee
where $\epsilon$ and its derivatives are evaluated at $\pi +k/2$. 
We see that the effective mass for the centre of mass motion 
is given by:
\be {1\over 2m}=\epsilon^{(2)}(\pi +k/2).\ee
$(1/m)$ vanishes at the inflection point, $k\to 2 k_{in}-2\pi\approx 
0.262\pi$ 
and the effective kinetic energy becomes quartic. At 
$k_{in}$, the coefficient of the quartic term is
$\epsilon^{(4)}/12\approx 3.341>0$. For a quartic 
kinetic energy, an arbitrarily weak attraction leads to 
a bound state, in both symmetric {\it and} antisymmetric channel. 
(For the antisymmetric case this can be seen by considering 
a trial wave-function of $\propto xe^{-x^2/(2w^2)}$. 
The potential energy can be assumed to be everywhere 
less than a square well of depth $v_0$ and width $a$. 
For $w\gg a$, the potential energy is less than 
a quantity $\propto -v_0(a/w)^3$, 
while the kinetic energy is $\propto (1/w)^4$. For 
small $v_0$, this has a negative minimum at $w\propto 1/v_0\gg a$, 
proving the existence of an antisymmetric bound state with 
binding energy $\propto v_0^4$.)
This argument implies that $k_c<2|\pi -k_{in}|\approx .262\pi$, 
since for any attractive potential, as the effective mass 
diverges the bound state will eventually form. It is 
interesting to note that our DMRG estimate of $k_c\approx 0.23\pi-0.24\pi$ 
is only very slightly less than $2|\pi -k_{in}|$, suggesting 
that the attractive interaction between magnons is weak. 
In addition, the splitting of the peak from the two magnon continuum varies as $(m - m_0)^2$ (where $m_0$ is the
mass at $k=k_c$) and thus
as $(k-k_c)^2$ within this bound state picture.

\begin{figure}[htbp]
\begin{center}
\includegraphics*[width=0.7\hsize,scale=1.0]{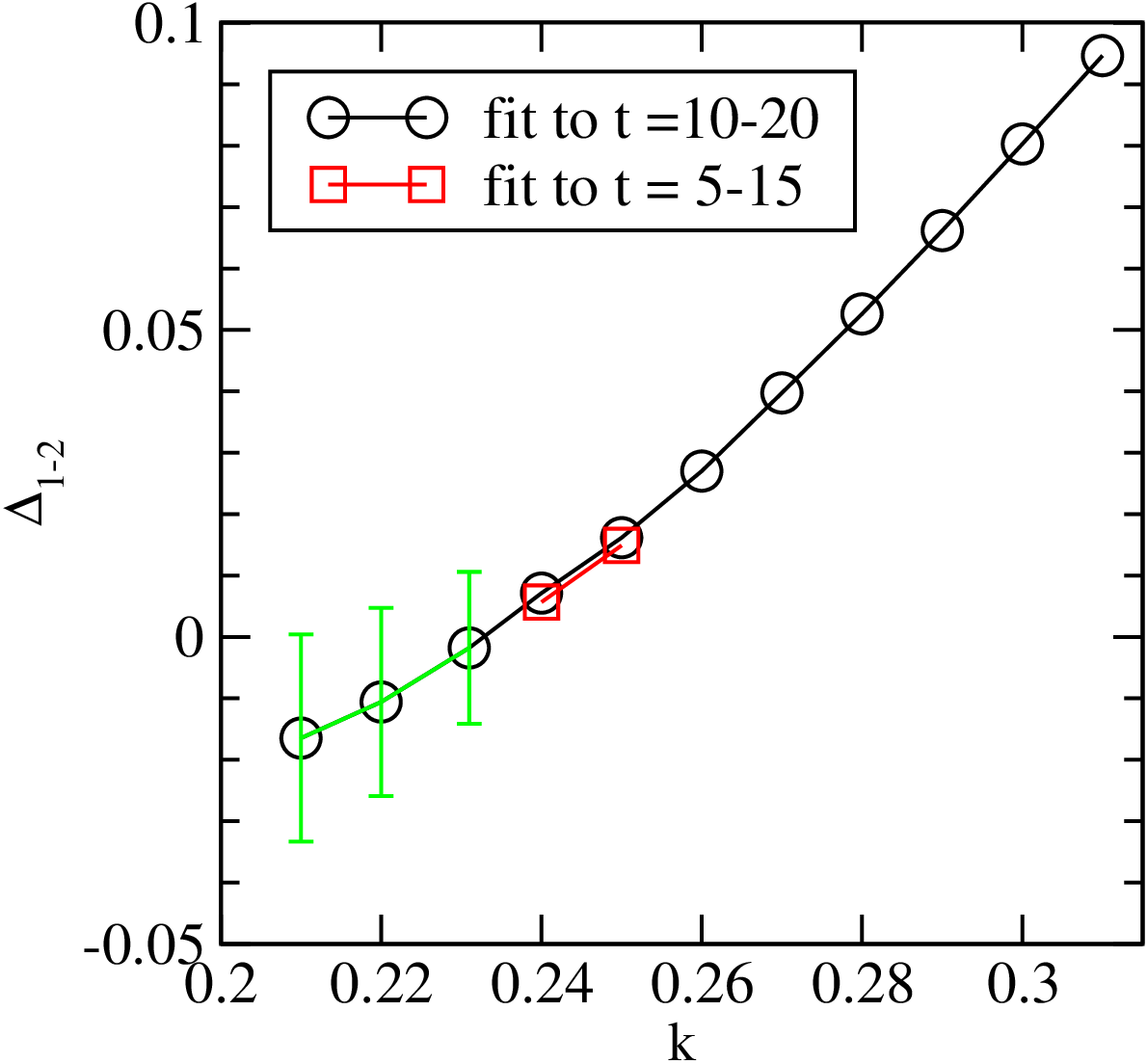}
\caption{Splitting between the single magnon peak and the bottom of the two magnon band
near $k_c$ determined from a fit of the time dependent DMRG data. For $\omega \ge 0.24$, the data was fit well
assuming the single magnon peak was a delta function; below a  better fit was obtained assuming  the peaks was
broadened as a Gaussian. The ``error bars'' indicate the width of the Gaussian. }
\label{splitnearkc}
\end{center}
\end{figure}

Now consider the second picture of the split-off, that of a magnon entering the continuum, but surviving
in a broadened form near $k_c$. In this case one would expect for $k>k_c$ the single magnon peak and the two
magnon continuum would vary independently with $k$ and that the splitting of the peak from
the two magnon continuum would be linear in $k-k_c$. In Fig. \ref{splitnearkc}, we show this splitting near $k_c$.
Indeed, the splitting appears to be linear in $k-k_c$.  Below $k_c$, the broadening grows very rapidly. If the
splitting is quadratic in $k-k_c$, it must be so only very close to $k_c$.

The behavior of the splitting versus $k-k_c$ favors the second picture of the split-off.  However, we cannot
rule out that the two-magnon bound state picture applies very close to $k_c$. 
Note that
within the
two magnon bound state picture, there does not seem to be a compelling reason for more than
a small fraction of the spectral weight to appear in the bound state. In fact, as shown in Fig. \ref{fratio},
about 50\% of the
spectral weight appears in the single magnon peak near $k\approx 0.25$, but the weight in the peak
is rapidly falling as $k$ is decreased.

\begin{figure}[htbp]
\begin{center}
\includegraphics*[width=0.7\hsize,scale=1.0]{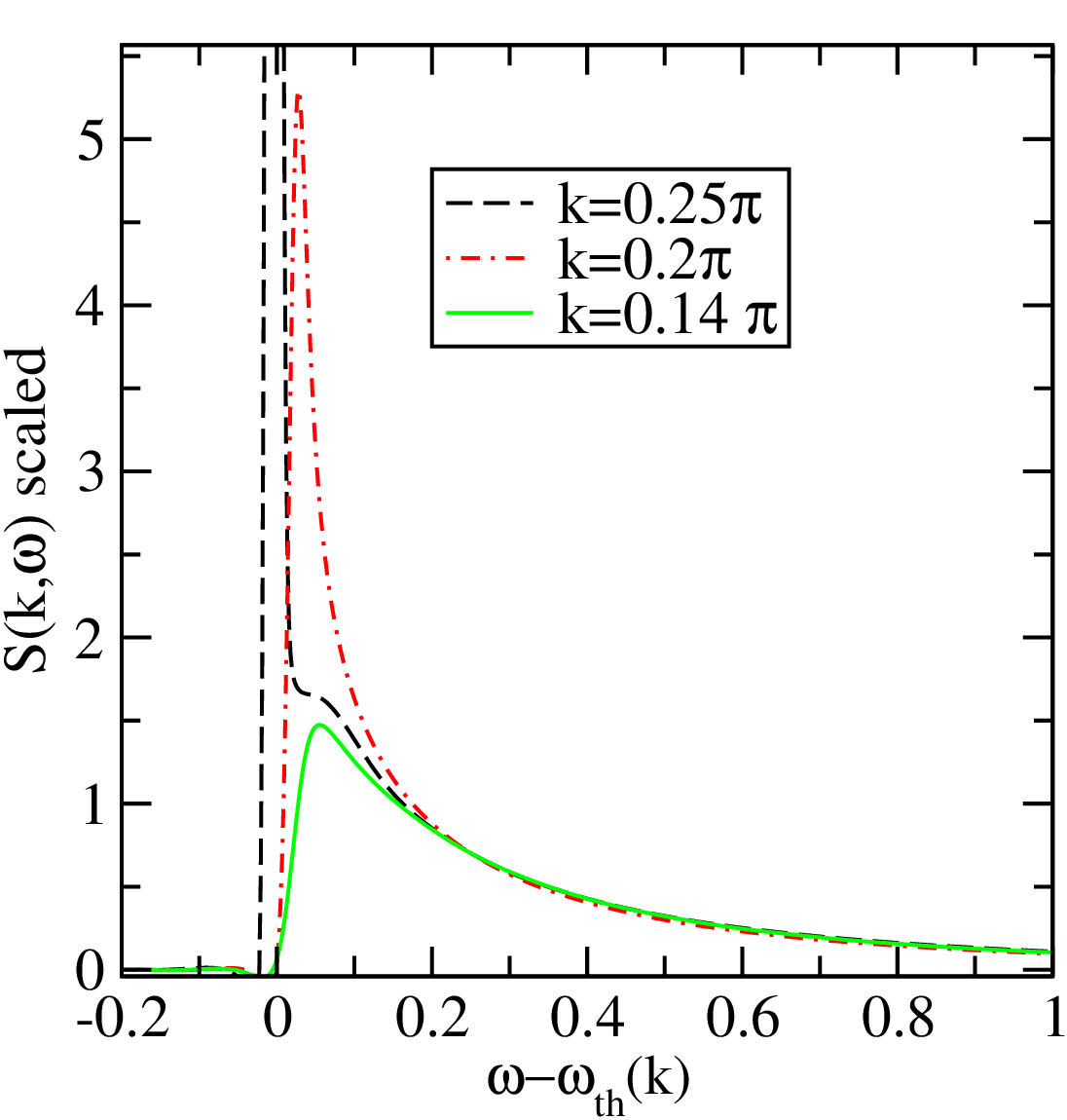}
\caption{$S(k,\omega)$ near and below $k_c \approx 0.23\pi-0.24 \pi$ for the 2nd order run. Each curve is shifted
by the two-magnon threshold energy $2 \varepsilon(k/2)$, and scaled by an arbitrary factor to make 
$S(k,\omega-\omega_{\rm th}=0.25)$
identical in each of the three curves.}
\label{Sbelowkc}
\end{center}
\end{figure}

It is also interesting to examine the line shape for $k$ 
slightly less than $k_c$.  In order to compare line shapes for different $k$'s, in Fig. \ref{Sbelowkc} we have
shifted the curves to make the two-magnon thresholds identical, and have scaled them to make them identical at
the  arbitrary point 
$\omega_{\rm th}+0.25$.  These curves were made using the linear prediction method. The curve for $k=0.2\pi$
shows a sharp resonance persists below but near $k_c$.
This resonance disappears by the time
$k$ is reduced to $0.14 \pi$.

\section{Conclusions}

The combination of time dependent DMRG and extrapolation of the time dependent correlation functions has
proved to be an extremely effective method for calculating spectral functions for the $S=1$ chain. We have been
able to study fine details of the spectra with much  greater resolution and accuracy than with any
previous method.  In comparing with free boson  and nonlinear sigma model predictions for features of
the spectra near $k=0$ and $k=\pi$, we find good qualitative agreement, but quantitative disagreements 
in the overall magnitude of the spectrum and in the high frequency tails.  Our results near $k_c$ where the single
magnon peak enters the two magnon continuum are better described in terms of a single magnon exhibiting
decay and  scattering
below $k_c$ rather than viewing the single magnon peak as the formation of a two magnon bound state above
$k_c$.

We acknowledge very helpful discussions with David Huse.
We acknowledge  support from the NSF under grant DMR-0605444 (SRW), from 
NSERC (IA),  and from CIfAR (IA).

\end{document}